\documentclass{svjour3}                     % onecolumn (standard format)
\smartqed  % flush right qed marks, e.g. at end of proof

\usepackage{booktabs}
\usepackage{amsmath,graphicx,centernot}
\newcommand{\CI}{\mathrel{\perp\mspace{-10mu}\perp}}
\newcommand{\nCI}{\centernot{\CI}}

\newcommand{\lt}{\ensuremath <}
\newcommand{\gt}{\ensuremath >}

\usepackage[colorinlistoftodos]{todonotes}

\newcommand{\change}[1]{\textcolor{red}{{#1}}}

\usepackage{hyperref}%
\usepackage{blindtext}%

\usepackage[ruled,vlined,linesnumbered]{algorithm2e}
\SetKwInput{Input}{Input}
\SetKwInput{Output}{Output}
\SetKw{To}{ to }
\SetKw{Print}{Print }

\newcommand{\RQone}{What does the literature proclaim about CI and software quality?}
\newcommand{\RQtwo}{Is the causal effect of CI on software quality empirically observable?}
\newcommand{\RQthree}{What would be an accurate causal theory for CI?}
\newcommand{\replication}{https://github.com/elieziosoares/ci\_quality\_study\_replication}

\usepackage{float}
\usepackage{cite}
\usepackage{amsmath,amssymb,amsfonts}
\usepackage{algorithmic}
\usepackage{graphicx}
\usepackage{textcomp}
\usepackage{xcolor}
\def\BibTeX{{\rm B\kern-.05em{\sc i\kern-.025em b}\kern-.08em
  T\kern-.1667em\lower.7ex\hbox{E}\kern-.125emX}}

\usepackage{tcolorbox}
\usepackage[linesnumbered,ruled,vlined]{algorithm2e}
\begin{document}

\title{Continuous Integration and Software Quality: A Causal Explanatory Study with Travis CI Projects}
%\title{Continuous Integration and Software Quality: A Causal Explanatory Study}

%\title{Conference Paper Title*\\

\author{Eliezio Soares         \and
        Daniel Alencar da Costa        \and
        Uirá Kulesza
}


\institute{Federal Institute of Rio Grande do Norte - IFRN \\
            Federal University of Rio Grande do Norte - UFRN \at
            Natal-RN, Brasil\\
              \email{eliezio.soares@ifrn.edu.br}             %\\
             %\emph{eliezio.soares@ifrn.edu.br} %of F. Author  %  if needed   
          \and
          University of Otago \at
          Dunedin, New Zealand   \\
          \email{danielcalencar@otago.ac.nz}      
          \and
          Federal University of Rio Grande do Norte - UFRN \at
          Natal-RN, Brasil\\
          \email{uira@dimap.ufrn.br}    
}

\date{Received: date / Accepted: date}

\maketitle

\begin{abstract}
	Continuous Integration (CI) is a software engineering practice that aims to reduce the cost and risk of code integration among teams. Recent empirical studies have confirmed associations between CI and the software quality (SQ). However, no existing study investigates causal relationships between CI and SQ. This paper investigates it by applying the causal Direct Acyclic Graphs (DAGs) technique. We combine two other strategies to support this technique: a literature review and a Mining Software Repository (MSR) study. 

    In the first stage, we review the literature to discover existing associations between CI and SQ, which help us create a ``literature-based causal DAG'' in the second stage. This DAG encapsulates the literature assumptions regarding CI and its influence on SQ. In the third stage, we analyze 12 activity months for 148 open-source projects by mining software repositories--- 74 CI and 74 no-CI projects. This MSR study is not a typical ``correlation is not causation'' study because it is used to verify the relationships uncovered in the causal DAG produced in the first stages. The fourth stage consists of testing the statistical implications from the ``literature-based causal DAG'' on our dataset. Finally, in the fifth stage, we build a DAG with observations from the literature and the dataset, the ``literature-date DAG''.

    In addition to the direct causal effect of CI on SQ, we find evidence of indirect effects of CI. For example, CI affects teams' communication, which positively impacts SQ. We also highlight the confounding effect of project age.
\end{abstract}

\section{Introduction \label{intro}}
Continuous integration (CI) is an increasingly popular software engineering practice used in a variety of domains and organizations~\cite{beck2010xp,fowler_ci,duvall2013continuous,bosch2014modeling}. CI proposes to reduce the costs and risks related to code integration by having a central daily code integration task. Code should be integrated every couple of hours while sharing the project knowledge, the code responsibility and reducing the risks related to late integration~\cite{beck2010xp,fowler_ci,duvall2013continuous,bosch2014modeling}.

Practitioners and existing literature have proclaimed several benefits related to using CI. CI is associated with improvements in testing practices, better quality assurance, a reduction in issue reports, among others~\cite{soares2021effects}. Although studies investigating associations are valuable, causal studies are important for decision making and deeper insights in an area of knowledge. To the best of our knowledge, there is a lack of studies that investigate causal relationships between CI and software quality.

Causal knowledge is valuable because it empowers stakeholders to make better decisions in software development~\cite{pearl2000models}, such as adopting CI or defining a communication policy. Our work investigates the potential causal relationship between CI and software quality to better understand the influence of CI on delivering better software products. To study the causal relationship between CI and software quality, we use an approach that consists of five interconnected stages. Each stage is guided by a research question described in the following.
%For this purpose, we employ three complementary methods: (i) first, a literature review to explore the relevant assumptions regarding CI, software quality, and adjacent relationships; (ii) second, we use Casual DAGs - a graphical-statistical technique - to enable us to draw domain assumptions and infer causal conclusions; and (iii) third, we conduct a mining software repository (MSR) study to build a dataset supporting the analysis of the statistical implications of the DAG on a real software projects dataset. 

\textbf{RQ1. \RQone}
To understand the variables that can potentially play a role in the relationship between CI and software quality, we conduct a literature review, which helps us define a causal DAG (i.e., a graphical-statistical technique - to enable us to draw domain assumptions and infer causal conclusions in a later stage~\cite{hernan2010}). The goal of the review is to discover the existing connections between CI and software quality, as well as marginal associations (e.g., associations between code smells and software quality), which will help us to draw a comprehensive DAG containing the existing studies' assumptions about how CI may influence software quality (and the potential confounding variables surrounding both CI and software quality). To study this relationship, we consider bug reports a proxy for software quality, similar to a previous study from Vasilescu et al.~\cite{vasilescu2015quality}.

\textbf{RQ2. \RQtwo}
Once we have a causal DAG containing a sufficient set of variables to analyze the relationship between $Continuous Integration$ and $Bug Report$, we can apply the d-Separation rules~\cite{pearl1995causal} and evaluate the raised set of testable statistical implications from the DAG built in RQ1. d-Separation is a set of graphical rules to identify if an association path exists between two or more variables in the DAG. From such associations, testable implications arise (d-Separation will be explained in Section~\ref{d-separation}). Next, we mine software repositories to collect observational data on the variables of the DAG and perform (un)conditional independence tests on our dataset to answer RQ2. Note that this differs from an ordinal ``causation is not correlation'' MSR study, as these tests are guided by the d-Separation rules derived from the causal DAG.

\textbf{RQ3. \RQthree}
Considering the investigations in RQ1 and RQ2 and the testable implications from the causal DAG in the dataset, we can analyze the hypotheses that failed in our analyses (i.e., statistical implications from the supposed relationships between the variables that were not supported by the data) and propose a new causal DAG. We do so, again, using the literature knowledge and empirical data to answer RQ3 but now considering the new and corrected causal DAG this time.

Our study innovates in offering the seminal empirical causal conclusions regarding CI and software quality (using the number of bug reports as a proxy for quality). This innovation helps evolve the current empirical knowledge of Continuous Integration as most of our empirical observations rely on statistical correlations instead of causal relationships.
We also contribute with research and practical implications for the software engineering community. Understanding the potential effects and which variables play relevant causal influences allows us to make beneficial software engineering decisions, e.g., seeking to maximize or minimize these effects, such as the influence of team communication decreasing the number of bug reports.

The rest of the paper is organized as follows: Section~\ref{motivation} presents the current state of knowledge of Continuous Integration and the need for causal studies. Additionally, Section~\ref{motivation} offers a technical background to understand the employed methodology. Section~\ref{method} explains the methodology to answer each research question. Section~\ref{results} presents the results of our studies for each research question. Section~\ref{discussion} discusses our results and findings. In Section~\ref{threats}, we present threats to our study's validity. Finally, Section~\ref{conclusion} concludes this study, highlighting the main findings.

\section{MOTIVATION AND RELATED WORK \label{motivation}}
Our study investigates potential causal relationships between continuous integration (CI) and software quality. This study is important because understanding causal effects can help practitioners to measure the actual benefits of CI. Given the predominance of statistical correlation studies, this work extends the current CI knowledge by offering causal conclusions. Our study helps researchers to remain aware of possible confounders, related variables, and adjacent associations when investigating the relationship between CI and software quality.

Through a systematic literature review (SLR), Soares et al.~\cite{soares2021effects} investigated the influence of CI on software development. Their study compiles most of the existing (non-causal) associations between CI and software quality. Based on current findings, the SLR concludes that CI may improve the time to develop and merge addressed issues as well as to reduce the number of reported defects~\cite{zhao2017impact,rahman2018characterizing,soares2021effects}. However, the SLR also identifies several CI associations that need further investigation. For example, the relationship between the development environment reliability and developers’ (over)confidence is due to trusting CI outcomes.

Despite all the effort invested in studying the potential benefits of CI~\cite{soares2021effects,bosch2014modeling,zhao2017impact,vasilescu2015quality,keskin2019software,pinto2018work,cassee2020silent},
the software engineering community still needs to benefit from a step further, which is to investigate causal relationships in existing studies. The difference between association and causation is critical. For example, inferring causal conclusions from associations may be harmful because of spurious associations, leading to false conclusions (e.g., confounding effect)~\cite{pearl2000models,greenland1999}. Unconsidered relationships among variables may confound the causal assumptions about the studied phenomenon. For example, one can observe that streets have puddles and people wear raincoats whenever it rains. However, if one assumes that puddles cause people to wear raincoats, one would fail to consider a relationship of a third variable (\textit{rain}) that causes both puddles and people wearing raincoats. In the following subsections, we explore this in more detail.

%For this purpose, Pearl \cite{pearl2000models} proposes the employment of a causal modeling framework based on causal Directed Acyclic Graphs (causal DAGs) to investigate and infer causation.

\subsection{Common Cause Principle \label{commonCause}} 
\begin{figure}[htbp]
	\centerline{\includegraphics[scale=.6]{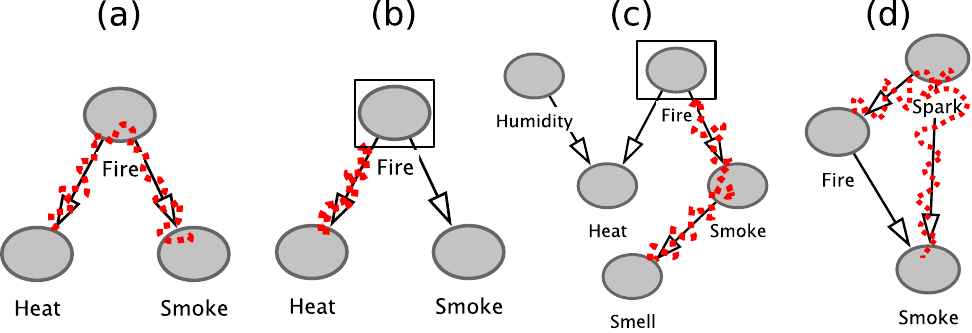}}
	\caption{In this picture, the ellipses are variables, the edges represent a relationship between variables, and the red dots above the edges represent an ``association flowing'' between variables. (a) Fire is the common cause of Heat and Smoke. Heat and Smoke are associated through Fire, i.e., the ``association flows'' between Heat and Smoke through Fire. (b) Conditioning on Fire, the association flow between Heat and Smoke is blocked. (c) Conditioning on Fire, Smoke becomes associated only with its descendent Smell. (d) Spark is a common cause of Fire and Smoke, opening a \textit{backdoor path} between these variables. Spark is a source of confounding.}
	\label{fig:dag-example}
\end{figure}

Reichenbach's Common Cause Principle~\cite{reichenbach1991direction} states that given two statistically dependent variables $X$ and $Y$, if one is not a cause of the other, then they may share a common cause $Z$, as shown in Fig.~\ref{fig:dag-example}(a). Conditioning on the common cause $Z$, then $X$ and $Y$ become independent. For example, Fig.~\ref{fig:dag-example}(a) shows an association ``flowing'' (the red dots above on the edges represent an ``association flow'') between $Heat$ and $Smoke$. They are associated because they share a common cause, which is $Fire$. Fig.~\ref{fig:dag-example}(b) shows the interrupted flow when conditioning on $Fire$, i.e., in the absence of $Fire$ (i.e., $Fire$=0), there is no association between $Heat$ and $Smoke$.

Therefore, if the famous adage states that ``correlation does not imply causation'' (i.e., statistical associations are not sufficient to determine causal relationships), on the other hand, ``there is no causation without association.'' The common cause principle establishes a relationship between statistical properties (i.e., association) and causal structures~\cite{peters2017elements}. In this way, it is possible to infer the existence of causal links from statistical dependencies (i.e., functional relationships between the variables) \cite{peters2017elements}. To infer causation, Pearl \cite{pearl2000models} proposed employing a causal modeling framework based on causal Directed Acyclic Graphs (causal DAGs). In the following subsections, we explain the theory proposed by Pearl because we apply his proposed theory in our study.

\subsection{Causal Directed Acyclic Graphs (Causal DAGs) \label{causalDAGs}}
%\textbf{Causal Directed Acyclic Graphs (Causal DAGs): } 
Pearl \cite{pearl2000models} argues that nature possesses causal mechanisms that, if described in detail, are deterministic functional relationships between variables. Some of these variables are unobservable, e.g., sometimes we see the smoke causing a fire alarm to be activated, but we can not see the fire. However, it is the fire that causes the existence of smoke. Pearl describes the causal discovery task as an induction game that contributes to identifying (from available observations or interventions) the organization of the mechanisms in the form of an acyclic causal structure~\cite{pearl2000models}. This causal structure is called a Directed Acyclic Graph (DAG). A DAG has: (i) nodes - that are variables with directed edges and no directed cycles; and (ii) edges that represent functional relationships between variables (see Fig.~\ref{fig:dag-example}(a))~\cite{pearl2000models,spirtes2000causation,hernan2010}.

Pearl \cite{pearl2000models} argues that nature possesses causal mechanisms that, if described in detail, are deterministic functional relationships between variables. Some of these variables are unobservable, e.g., sometimes we see the smoke causing a fire alarm to be activated, but we can not see the fire. However, it is the fire that causes the existence of smoke. Pearl describes the causal discovery task as an induction game that contributes to identifying (from available observations or interventions) the organization of the mechanisms in the form of an acyclic causal structure~\cite{pearl2000models}. This causal structure is called a Directed Acyclic Graph (DAG). A DAG has: (i) nodes - that are variables with directed edges and no directed cycles; and (ii) edges that represent functional relationships between variables (see Fig.~\ref{fig:dag-example}(a))~\cite{pearl2000models,spirtes2000causation,hernan2010}.

In turn, Pearl and Verma \cite{pearl1991theory} defined a Causal Markov Condition stating that a DAG should have a node distribution \(\displaystyle N=\lbrace N_{1},...,N_{n}\rbrace \) such that, for each j, \(\displaystyle N_{j} \) is independent of its non-descendants conditioning on its parents \cite{pearl2000models,hernan2010,spirtes2000causation}. Considering the example of Fig.~\ref{fig:dag-example}(c), the Markov Condition implies that, if conditioning on the parent $Fire$, $Smoke$ becomes independent of all other variables on the DAG, except its descendent $Smell$. That means that $Smoke$ exists ($Smoke=1$), but there is not necessarilly $Heat$, because we condition on $Fire$ ($Fire=0$).

Reichenbach's Common Cause Principle (see section~\ref{commonCause}) and the Markov Condition imply that a causal DAG should contain the common causes of any pair of variables \cite{pearl2000models,hernan2010}. Thus, to build a sufficient causal DAG of a phenomenon, it is essential to know the significant common causes among the variables involved in that phenomenon, i.e., if two variables $Heat$ and $Smoke$ in the DAG share a common cause $Fire$, then $Fire$ should be represented in the DAG, as shown in Fig.~\ref{fig:dag-example}(c). Since this is a recursive criterion, in Fig.~\ref{fig:dag-example}(d), $Spark$ should be present since it is a common cause of $Fire$ and  $Smoke$.

Discovering causal structures to build a DAG that correctly describes a phenomenon, i.e., the set of variables a DAG should contain and the relationships between variables, is challenging. There are at least three strategies to obtain a DAG: prior knowledge, guessing-and-testing, or discovery algorithms ~\cite{shalizi2021}. Prior knowledge is a source to build causal DAGs, but since there is a link between causal structures and statistical properties (see section~\ref{commonCause}), it is possible to test the correctness of a built DAG if we have access to observational data. Section~\ref{d-separation} details the statistical properties and their testable implications.

\subsection{d-Separation and the Testable Implications of the DAGs \label{d-separation}}
%\textbf{d-Separation and the Testable Implications of the DAGs: \label{d-separation}}
A helpful approach to visually understand DAGs is to assume that associations ``flow'' through the edges of the DAG~\cite{hernan2010}. Using the illustration from Fig.~\ref{fig:dag-example}(a), the association flows freely between $Heat$ and $Smoke$. However, intervening on the values of $Fire$ (as in the Fig.~\ref{fig:dag-example}(b)), this flow may be interrupted since it represents a common cause for $Heat$ and $Smoke$~\cite{hernan2010}. A set of graphical rules, called d-separation, were formalized by Pearl~\cite{pearl1995causal} to infer associational conclusions from causal DAGs.

D-separation is a set of graphical rules to define if a path in the DAG is blocked or open. To understand the rules, we consider three structural patterns: (i) Chain:  $Fire\to Smoke\to Smell$ (see Fig.~\ref{fig:dag-example}(c)); (ii) Fork:  $Heat\leftarrow Fire\to Smoke$ (see Fig.~\ref{fig:dag-example}(a)); and (iii) Collider: $Humidity\to Heat\leftarrow Fire$ (see Fig.~\ref{fig:dag-example}(c)). 
In the chain, $Fire$ and $Smell$ are dependent ($Fire \nCI Smell$) but become independent (i.e., blocked), conditioning on $Smoke$ ($Fire \CI Smell \mid Smoke$). In the fork, $Heat$ and $Smoke$ are dependent ($Heat \nCI Smoke$) unless we condition on $Fire$ ($Heat \CI Smoke \mid Fire$). In the collider, $Humidity$ and $Fire$ are independent ($Humidity \CI Fire$)  because a collider blocks the association flow. Conditioning on the collider (or one of its descendants) opens that flow, making $Humidity$ and $Fire$ dependent(e.g., $Humidity \nCI Fire \mid Heat$) . We provide a summary below:

\begin{itemize}
    \item Chain: $X\to Z\to Y$  $\Rightarrow$ $X \nCI Y$ and $X \CI Y \mid Z$
    \item Fork: $X\leftarrow Z\to Y$  $\Rightarrow$ $X \nCI Y$ and $X \CI Y \mid Z$
    \item Collider: $X\to Y\leftarrow Z$  $\Rightarrow$ $X \CI Z$ and $X \nCI Z \mid Y$
\end{itemize}

Based on these d-Separation rules, all DAGs have a consequent set of testable statistical implications. These statistical implications inferred from the graphical analysis allow us to verify if the structure of a DAG is consistent with an empirical dataset through conditional independence tests on the data. Pearl~\cite{pearl2000models} defines a causal model as a pair $M=\lt D,\Theta _D\gt$ consisting of a causal structure $D$ and a set of parameters $\Theta _D$ compatible with D. Thus, relying on d-Separation rules and a representative dataset, the causal DAG technique also allows building a causal structure through both guessing-and-testing as well as discovery algorithms~\cite{shalizi2021}.

This study aims to understand how Continuous Integration affects Software Quality. We use a combination of two approaches: (i) prior knowledge to draw the existing assumptions; and (ii) guessing-and-testing to obtain a final causal structure. To introduce more rigor, we consider prior knowledge by gathering it from the existing literature and drawing an initial DAG with assumptions regarding the influence of CI on software quality. Afterward, we refine our understanding by applying guessing-and-testing that relies on the statistical implications that we derive from the initial DAG and their validations by using our dataset, i.e., we collect data by mining software repositories. Section~\ref{method} details these methodological steps. 

\subsection{Backdoor Paths and Confounding \label{confounding}}
A common source of bias is due to common causes between two variables, such as the presence of a cause $Spark$ shared by the treatment $Fire$ and the outcome $Smoke$ (see Fig.~\ref{fig:dag-example}(d)). This common cause results in a path between $Fire$ and $Smoke$ that is not a direct edge. The path is through the back door, i.e., $Fire \leftarrow Spark \to Smoke$ generating another association flow between the treatment $Fire$ and the outcome $Smoke$. We refer to this kind of structure as a \textit{backdoor} path, and the bias caused by the backdoor path as \textit{confounding}.

Causal DAGs have grown in popularity in several fields (econometrics, epidemiology, and climate science, among others). Greenland, Pearl, and Robins~\cite{greenland1999causal} present causal DAGs as a tool for identifying variables that must be measured and controlled to obtain unconfounded causal effect estimates for epidemiologic research. Shmueli~\cite{shmueli2010explain} sees causal DAGs as a common causal inference method for testing causal hypotheses on observational data. 

\section{METHODOLOGY \label{method}}

\begin{figure*}[htbp]
    \centerline{\includegraphics[scale=.35]{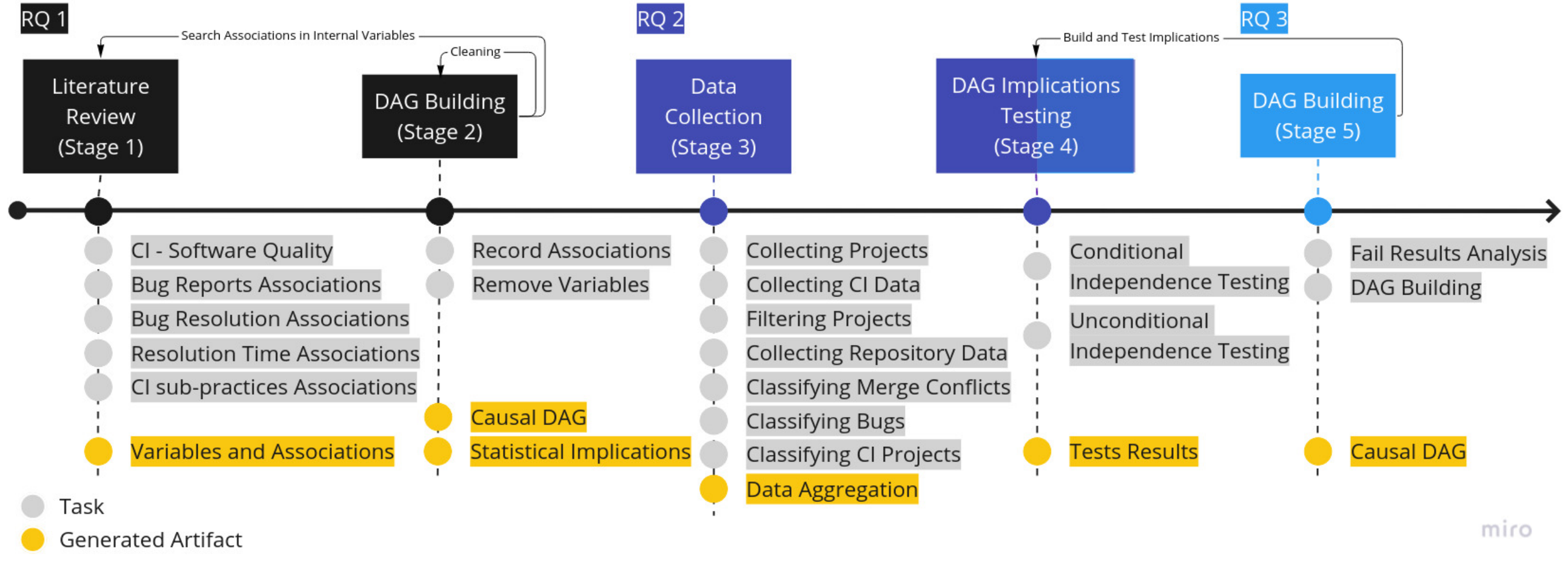}}
    \caption{Research method pipeline.}
    \label{fig:pipeline}
  \end{figure*}

We follow a pipeline of 5 stages shown in Fig.~\ref{fig:pipeline} to answer our research questions: The Literature Review (stage 1) and DAG Building (stage 2) stages contribute to answering RQ1. The data collection from repositories (stage 3) and the DAG Implications Testing (stage 4) stages address RQ2. Lastly, another stage of DAG Building (stage 5) performed in an iterative manner with the DAG Implications Testing stage (stage 4) until a final DAG is obtained containing literature and data consistency, answers RQ3. Next, we detail the different stages of our research methodology.

%\begin{comment}
  %\end{comment}

\subsection{\textbf{\RQone} \label{subrq1}}
As depicted in Fig.~\ref{fig:pipeline}, we perform two research stages to answer RQ1. First, the literature review compiles the existing assumptions about how CI may impact software quality and allow us to map variables, associations, and every common cause (i.e., the $Fire$ in the example of Fig.~\ref{fig:dag-example}(b)) for any two or more variables in the DAG. This mapped information is input to stage 2 - DAG building. We perform these two stages iteratively to build a causal DAG representing the most up-to-date literature knowledge.

As discussed in section~\ref{motivation}, prior knowledge is an existing approach to building causal structures regarding a phenomenon. Then in this first research question, we intend to draw a causal DAG based on the literature assumptions regarding continuous integration (CI) and software quality. Although most of the studies in software engineering are association studies, we discussed the relationship between association and causation in section~\ref{motivation} and how d-Separation rules raise testable statistical implications from causal DAGs~\cite{pearl2000models}. Thus, we begin by mapping the existing knowledge to infer causality, even if they come from association studies.

\subsubsection{\textbf{Literature Review\label{section-literature-survey}}}

To review the relationship between continuous integration (CI) and software quality (SQ), we consider ``issues'', ``bugs'', and/or ``defects'' as proxies of software quality. Since we are also interested in knowing the common causes (i.e., the $Fire$ in Fig.~\ref{fig:dag-example}(a)) between CI and these variables (i.e., `issues'', ``bugs'', or ``defects''), we review the literature for each one of these variables.

We also review the literature for each new variable found during the DAG building in an iterative manner. We perform the searches on Google Scholar\footnote{https://scholar.google.com/} using the name of variables or synonyms. To cover a broader range of studies and mitigate search bias, we prioritize systematic literature reviews on the search. For instance, we search for software bug relationships using the search string ``systematic literature review software bugs''. This search has a broader perspective and not necessarily the studies mention CI, but it is essential to detect correlated variables and their relationships. 

In addition, we search the literature for relationships among the following CI sub-practices --- tests practices, integration frequency, build health maintenance, and quick fixes of broken builds~\cite{beck2010xp,duvall2013continuous,fowler_ci}. In the full search, we selected 39 studies, of which 12 are systematic literature reviews. All the selected studies are referenced, and a complete list is available in our replication package\footnote{\replication}. We believe our literature review fits our goal of finding the claims surrounding CI and SQ, especially because of the systematic literature reviews, as they have already systematically compiled a comprehensive view of the areas of CI and SQ.

%This phase includes all variables and associations in the DAG and the possible inner relationships.

\begin{table}[H]
  \caption{Connections identified in the literature about CI and software quality variables.}
  \label{tab:ci-bug-associations}
  \begin{tabular}{|l|l|}
  \hline
  \textbf{Connection} & \textbf{Rationale} \\ \hline
  $CI\to Bug Resolution$                &   \begin{tabular}[c]{@{}l@{}}Projects presents more resolved issues and bugs after adoption\\of CI \cite{rahman2018characterizing}.\end{tabular}             \\ \hline
  $CI\to Resolution Time$              &   \begin{tabular}[c]{@{}l@{}}CI is related to an increasing in the number of issues closed by\\period, helping to spend less time debugging and more time\\adding features \cite{keskin2019software,zhao2017impact}.\end{tabular}             \\ \hline
  $CI\to Bug Report$                    &   \begin{tabular}[c]{@{}l@{}}CI teams discover more bugs than no-CI teams, and CI\\projects present fewer defects than no-CI projects \cite{amrit2018effectiveness,vasilescu2015quality}.\end{tabular}             \\ \hline
  $CI\to Transparency$                    &   \begin{tabular}[c]{@{}l@{}}CI is associated with a transparency increase, facilitating\\collaboration \cite{kerzazi2014why}.\end{tabular}             \\ \hline
  $CI\to Communication$                    &   \begin{tabular}[c]{@{}l@{}}The general discussion, the number of line-level review\\comments, and change-inducing review comments tend to\\decrease after CI adoption without affecting pull\\request activity \cite{cassee2020silent}.\end{tabular}             \\ \hline
  $CI\to Overconfidence$                  &   \begin{tabular}[c]{@{}l@{}}CI developers are reported as suffering from a false sense of\\confidence (when blindly trusting the tests) \cite{pinto2017inadequate,pinto2018work}.\end{tabular}             \\ \hline
  $CI\to Technical Challenges$            &   \begin{tabular}[c]{@{}l@{}}Configuring the build environment, the tools, and practices\\impose challenges for CI teams \cite{pinto2017inadequate,debbiche2014challenges}.\end{tabular}             \\ \hline
  $CI\to Tests Volume$            &   \begin{tabular}[c]{@{}l@{}}CI is associated with an increase in the test ratio \cite{sizilio2019empirical}.\end{tabular}             \\ \hline
  $CI\to Commit Frequency$            &   \begin{tabular}[c]{@{}l@{}}CI is linked to a change in the commits pattern \cite{zhao2017impact,rahman2018characterizing}.\end{tabular}             \\ \hline

  \end{tabular}
\end{table}
%\cite{keskin2019software,zhao2017impact,rahman2018characterizing,amrit2018effectiveness,vasilescu2015quality}

%The literature review is our foundation to find a model representing the literature assumptions on the relation between continuous integration and software quality, answering the RQ1 (see Section~\ref{rq1}). 
\textbf{Continuous Integration and Software Quality:}
Soares et al. \cite{soares2021effects} conducted a systematic literature review on the associations between CI and software development as a whole. The review highlights associations between CI and an increase in bug/issue resolution~\cite{rahman2018characterizing}. For this reason, our literature-based DAG starts from the connection between continuous integration and bug resolution. The notation $CI \to Bug Resolution$ represents a causal flow from CI to Bug Resolution. We expand our DAG based on other studies that raise other diverse CI associations (as summarized in Table~\ref{tab:ci-bug-associations}).

\textbf{Confounders related to bug reports:}
To investigate potential confounders to the effect of $CI\to Bug Report$, we search for other factors associated with Bug Rerport. Table~\ref{tab:bugreports-associations} shows the associations extracted from a taxonomy by Huang et al. \cite{huang2012taxonomy}, a systematic literature review by A. Cairo et al. \cite{cairo2018impact}, and a mixed mining software repositories (MSR)-survey study by Vasilescu et al. \cite{vasilescu2015quality}.

\begin{table}[H]
  \caption{Connections identified in the literature about bug reports.}
  \label{tab:bugreports-associations}
  \begin{tabular}{|l|l|}
    \hline
    \textbf{Connection} & \textbf{Rationale}                                                                                                                                                                               \\ \hline
    \begin{tabular}[c]{@{}l@{}}$Lack Of Knowledge$\\ $\to Bug Report$\end{tabular}                & \begin{tabular}[c]{@{}l@{}}Insufficient domain and linguistic knowledge are presented as\\possible human root causes for software defects \cite{huang2012taxonomy}.\end{tabular} \\ \hline
    \begin{tabular}[c]{@{}l@{}}$Lack TechKnowledge$\\ $\to Bug Report$\end{tabular}            & \begin{tabular}[c]{@{}l@{}}Insufficient programming and strategy knowledge and failure to catch\\the specific feature of the problems are mapped as possible human\\root causes for software defects  \cite{huang2012taxonomy}.\end{tabular} \\ \hline
    \begin{tabular}[c]{@{}l@{}}$Requirem Problem$\\ $\to Bug Report$\end{tabular}              & \begin{tabular}[c]{@{}l@{}}Requirement management problems and a misunderstanding of\\requirements and design specifications are reported as possible human\\causes of software defects \cite{huang2012taxonomy}.\end{tabular} \\ \hline
    \begin{tabular}[c]{@{}l@{}}$Overconfidence$\\ $\to Bug Report$\end{tabular}                  & \begin{tabular}[c]{@{}l@{}}Overconfidence and confirmation bias contributes to evaluation errors\\and software defects \cite{huang2012taxonomy}.\end{tabular} \\ \hline
    \begin{tabular}[c]{@{}l@{}}$Inattention\to$\\ $Bug Report$\end{tabular}                     & \begin{tabular}[c]{@{}l@{}}Interruptions and other kinds of inattention are reported as possible\\human causes of software defects \cite{huang2012taxonomy}.\end{tabular} \\ \hline
    \begin{tabular}[c]{@{}l@{}}$Communication$\\ $\to Bug Report$\end{tabular}                   & \begin{tabular}[c]{@{}l@{}}Communication problems lead to expression and comprehension\\errors \cite{huang2012taxonomy}.\end{tabular} \\ \hline
    \begin{tabular}[c]{@{}l@{}}$Config Management$\\ $\to Bug Report$\end{tabular}        & \begin{tabular}[c]{@{}l@{}}Configuration management problems lead to process errors \cite{huang2012taxonomy}.\end{tabular} \\ \hline
    \begin{tabular}[c]{@{}l@{}}$Tools\to$\\ $Bug Report$\end{tabular}                           & \begin{tabular}[c]{@{}l@{}}Tools problems like compiler induced defects are possible root causes\\of software defects \cite{huang2012taxonomy}.\end{tabular} \\ \hline
    \begin{tabular}[c]{@{}l@{}}$Code Smells\to$\\ $Bug Report$\end{tabular}        & \begin{tabular}[c]{@{}l@{}}Code smells on the occcurrence of bugs \cite{cairo2018impact}.\end{tabular} \\ \hline
    \begin{tabular}[c]{@{}l@{}}$Number Of Forks\to$\\ $Bug Report$\end{tabular}        & \begin{tabular}[c]{@{}l@{}}The number of forks has an association with an increase in bug\\reports \cite{vasilescu2015quality}.\end{tabular} \\ \hline
    %\begin{tabular}[c]{@{}l@{}}$Proj Volume\to$\\ $Bug Report$\end{tabular}        & \begin{tabular}[c]{@{}l@{}} \cite{vasilescu2015quality}.\end{tabular} \\ \hline
    \begin{tabular}[c]{@{}l@{}}$Proj Age\to$\\ $Bug Report$\end{tabular}            & \begin{tabular}[c]{@{}l@{}}Project age has a significant negative effect on the count of bugs\\reported by core developers \cite{vasilescu2015quality}.\end{tabular} \\ \hline
    \begin{tabular}[c]{@{}l@{}}$Proj Popularity\to$\\ $Bug Report$\end{tabular}     & \begin{tabular}[c]{@{}l@{}}Project’s popularity has a significant negative effect on the count\\of bugs reported by core developers \cite{vasilescu2015quality}.\end{tabular} \\ \hline
    \begin{tabular}[c]{@{}l@{}}$Quant Issues\to$\\ $Bug Report$\end{tabular}        & \begin{tabular}[c]{@{}l@{}}The number of non-bug issue reports has a significant and positive\\effect on the response \cite{vasilescu2015quality}.\end{tabular} \\ \hline
    \begin{tabular}[c]{@{}l@{}}$Tests Volume\to$\\ $Bug Report$\end{tabular}        & \begin{tabular}[c]{@{}l@{}}The size of test files has a negative effect on bug reports \cite{vasilescu2015quality}.\end{tabular} \\ \hline
  \end{tabular}
\end{table}

\textbf{Confounders related to bug resolution:}
To investigate potential confounders to the effect of $CI\to Bug Resolution$, we search for other factors associated with $Bug Resolution$ in the literature. We found that $Maintainability$, $Analysability$, $Changeability$, $Stability$, $Testability$, $Project Volume$, $Duplication$, $Unit size$, $Unit complexity$, and $Module coupling$ all share an association with $Bug Resolution$. For the sake of readability, we group all these relationships into $Internal Quality\to Bug Resolution$~\cite{bijlsma2012faster}. We also found the associations $Communication\to Bug Resolution$ and $Issue Priority\to Bug Resolution$ in the literature review from Zhang et al.~\cite{zhang2016literature}. Table~\ref{tab:bugresolution-associations} shows all these associations and their rationales.

\begin{table}[H]
  \caption{Connections identified in the literature about bug resolution.}
  \label{tab:bugresolution-associations}
  \begin{tabular}{|l|l|}
    \hline
    \textbf{Connection} & \textbf{Rationale}                                                                                                                                                                               \\ \hline
    \begin{tabular}[c]{@{}l@{}}$Internal Quality\to$\\ $Bug Resolution$\end{tabular}                & \begin{tabular}[c]{@{}l@{}}Elements of $Internal Quality$, such as maintainability, analysability,\\changeability, stability, testability, project volume, duplication, unit size,\\unit complexity, and module coupling, present significant correlation\\with defect resolution efficiency \cite{huang2012taxonomy}.\end{tabular} \\ \hline
    \begin{tabular}[c]{@{}l@{}}$Communication\to$\\ $Bug Resolution$\end{tabular}                   & \begin{tabular}[c]{@{}l@{}}Human and data elements such as comments, severity, product,\\component, among others, can improve the performance of bug\\resolution \cite{zhang2016literature}.\end{tabular} \\ \hline
    \begin{tabular}[c]{@{}l@{}}$Issue Priority\to$\\ $Bug Resolution$\end{tabular}                  & \begin{tabular}[c]{@{}l@{}}Priority and severity are non-textual factors of a bug report\\that enhance the capability of bug resolution \cite{zhang2016literature}.\end{tabular} \\ \hline
  \end{tabular}
\end{table}

\textbf{Confounders related to resolution time:}
Table~\ref{tab:resolutiontime-associations} shows associations between other factors than CI and $Resolution Time$. We found associations extracted from the literature including variables $Issue Type$, $Communication$ (e.g., comments in issues, bug reports, pull requests), $Issue Priority$, $Commit Frequency$, $Operate System$, and $Issue Description$.

\begin{table}[H]
  \caption{Connections identified in the literature about the resolution time.}
  \label{tab:resolutiontime-associations}
  \begin{tabular}{|l|l|}
    \hline
    \textbf{Connection} & \textbf{Rationale}                                                                                                                                                                               \\ \hline
    \begin{tabular}[c]{@{}l@{}}$Issue Type\to$\\ $Resolution Time$\end{tabular}                & \begin{tabular}[c]{@{}l@{}}Issue fixing times are different for different issue types \cite{murgia2014influence,licorish2017exploring,zhang2012empirical,mockus2000identifying}.\end{tabular} \\ \hline
    \begin{tabular}[c]{@{}l@{}}$Communication\to$\\ $Resolution Time$\end{tabular}             & \begin{tabular}[c]{@{}l@{}}The number of comments and the max length of all comments\\in the bug reports impact the resolution time. Bugs with little\\discussion tend to be resolved quickly~\cite{panjer2007predicting,zhang2012empirical}.\end{tabular} \\ \hline
    \begin{tabular}[c]{@{}l@{}}$Issue Priority\to$\\ $Resolution Time$\end{tabular}            & \begin{tabular}[c]{@{}l@{}}The severity of a bug report influences the delay before fixing it.\\As high the severity level, the fewer the delay \cite{zhang2012empirical}.\end{tabular} \\ \hline
    \begin{tabular}[c]{@{}l@{}}$Commit Size\to$\\ $Resolution Time$\end{tabular}               & \begin{tabular}[c]{@{}l@{}}The size of code churn (number of methods) impacts the delay\\before fixing a bug report \cite{zhang2012empirical}.\end{tabular} \\ \hline
    \begin{tabular}[c]{@{}l@{}}$Operate System\to$\\ $Resolution Time$\end{tabular}            & \begin{tabular}[c]{@{}l@{}}The median delay before fixing a bug found on Linux is shorter\\than other OS \cite{zhang2012empirical}.\end{tabular} \\ \hline
    \begin{tabular}[c]{@{}l@{}}$Issue Description\to$\\ $Resolution Time$\end{tabular}         & \begin{tabular}[c]{@{}l@{}}Increasing the literal length of the bug report description can\\increase delay until the team checks it as resolved \cite{zhang2012empirical}.\end{tabular} \\ \hline
  \end{tabular}
\end{table}

\textbf{Internal confounders:}
We also consider internal relationships between every discovered variable to understand possible confounding scenarios in our causal DAG. That means considering potential associations between peripheral variables, e.g.,  $Issue Type\to Commit Frequency$. Table~\ref{tab:internal-bugs-associations} shows such potential associations and their rationales. 

\begin{table}[H]
  \caption{Internal connections cataloged among the literature regarding the discovered variables.}
  \label{tab:internal-bugs-associations}
  \begin{center}
  \begin{tabular}{|l|l|}
    \hline
    \textbf{Connection} & \textbf{Rationale}                                                                                                                                                                               \\ \hline
    \begin{tabular}[c]{@{}l@{}}$Issue Type\to$\\ $Commit Size$\end{tabular}         & \begin{tabular}[c]{@{}l@{}}The issue type is associated with the size of the code churn \cite{hindle2008large}.\end{tabular} \\ \hline
    \begin{tabular}[c]{@{}l@{}}$Issue Type\to$\\ $Engagement$\end{tabular}          & \begin{tabular}[c]{@{}l@{}}Developers tend to spend more effort engaging with one another\\regarding new features and software extensions than in defects \cite{licorish2017exploring}.\end{tabular} \\ \hline
    \begin{tabular}[c]{@{}l@{}}$Issue Type\to$\\ $Info Sharing$\end{tabular}        & \begin{tabular}[c]{@{}l@{}}Developers tend to share more information on defects and\\enhancements than support tasks \cite{licorish2017exploring}.\end{tabular} \\ \hline
    \begin{tabular}[c]{@{}l@{}}$Issue Type\to$\\ $Communication$\end{tabular}       & \begin{tabular}[c]{@{}l@{}}A higher number of comments is associate with enhancements\\and defects \cite{licorish2017exploring}.\end{tabular} \\ \hline
    \begin{tabular}[c]{@{}l@{}}$Issue Type\to$\\ $Difficulty Level$\end{tabular}    & \begin{tabular}[c]{@{}l@{}}There is an association between the difficulty of a change\\and its type \cite{mockus2000identifying}.\end{tabular} \\ \hline
    \begin{tabular}[c]{@{}l@{}}$Stability\to$\\ $Technical Challenges$\end{tabular} & \begin{tabular}[c]{@{}l@{}}The maturity of the tools, infrastructure, and CI activities\\imposes challenges to practitioners \cite{debbiche2014challenges}. The stability and maturity\\of the software under test affect the maintenance effort of tests \cite{garousi2016when}.\end{tabular} \\ \hline
  \end{tabular}
\end{center}
\end{table}

\begin{figure}[htbp]
  \centerline{\includegraphics[scale=.4]{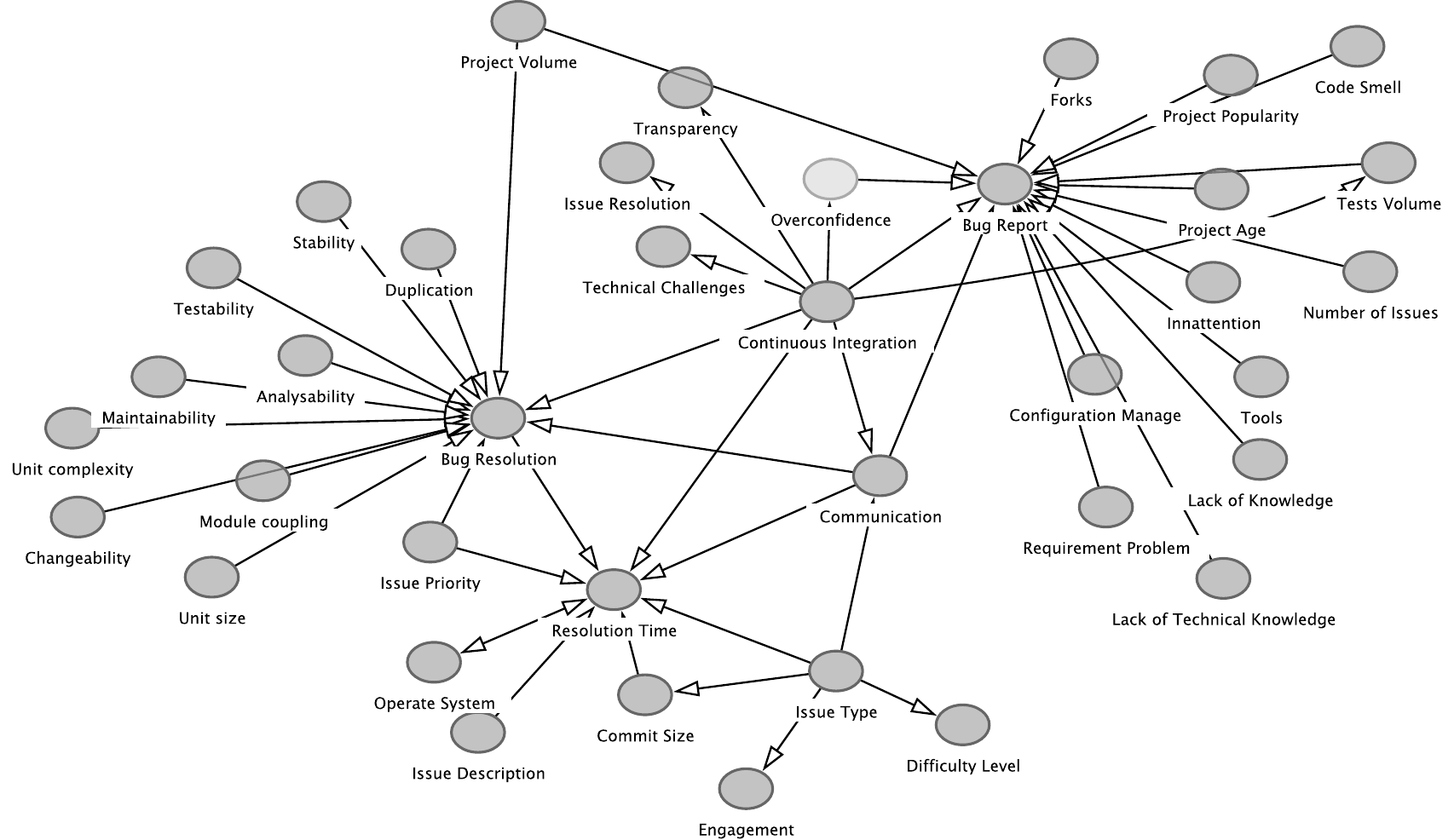}}
  \caption{Partial causal DAG for bug reports associations. }
  \label{fig:dag-bugs}
\end{figure}

With these previous connections, we build the partial causal DAG represented in Fig.~\ref{fig:dag-bugs}, where continuous integration is the intervention, and $Bug Resolution$ and $Bug Report$ are potential outcomes.

\textbf{Confounders related to Continuous Integration and its sub-practices:}
To investigate variables related to continuous integration (CI), we also consider its sub-practices~\cite{beck2010xp,duvall2013continuous,fowler_ci}. We consider the practices: automated tests practices, integration frequency, build health maintenance, and time to fix a broken build~\cite{duvall2013continuous}. We represent an association with a sub-practice of CI as an association with CI itself. For example, $Automated Tests\to Confidence$ will be represented as $Continuous Integration\to Confidence$. This decision avoids intangible discussions about what comes first, CI or Automated Tests, for instance. To implement CI, we consider that such practices are implicit, i.e., there is no CI practice without implementing those sub-practices~\cite{soares2021effects,duvall2013continuous}. Therefore, distinguishing the effect of CI and the effect of its sub-practices would be challenging. Table~\ref{tab:tests-associations} shows the associations relating to testing, while Table~\ref{tab:build-associations} groups and presents those relating build practices.

\begin{table}[H]
  \caption{CI connections cataloged among the literature from the perspective of test practices.}
  \label{tab:tests-associations}
  \begin{center}
  \begin{tabular}{|l|l|}
    \hline
    \textbf{Connection} & \textbf{Rationale}                                                                                                                                                                               \\ \hline
    \begin{tabular}[c]{@{}l@{}}$Automated Tests\to$\\ $Bug Report$\end{tabular}           & \begin{tabular}[c]{@{}l@{}}Automated tests are related to improved product quality in terms\\of fewer defects in the software \cite{dudekula2012benefits}.\end{tabular} \\ \hline
    \begin{tabular}[c]{@{}l@{}}$Automated Tests\to$\\ $Code Coverage$\end{tabular}             & \begin{tabular}[c]{@{}l@{}}Automated tests are related to high coverage of code \cite{dudekula2012benefits}.\end{tabular} \\ \hline
    \begin{tabular}[c]{@{}l@{}}$Automated Tests\to$\\ $Work Time$\end{tabular}            & \begin{tabular}[c]{@{}l@{}}Automated tests are related to reduced testing time \cite{dudekula2012benefits}.\end{tabular} \\ \hline
    \begin{tabular}[c]{@{}l@{}}$Automated Tests\to$\\ $Confidence$\end{tabular}           & \begin{tabular}[c]{@{}l@{}}Automated tests are related to increased confidence in the quality\\of the system \cite{dudekula2012benefits}.\end{tabular} \\ \hline
    \begin{tabular}[c]{@{}l@{}}$Automated Tests\to$\\ $Human Effort$\end{tabular}         & \begin{tabular}[c]{@{}l@{}}Automated tests are related to the less human effort that can be\\redirected for\\other activities \cite{dudekula2012benefits}.\end{tabular} \\ \hline
    \begin{tabular}[c]{@{}l@{}}$Automated Tests\to$\\ $Cost$\end{tabular}                 & \begin{tabular}[c]{@{}l@{}}Automated tests are related to a reduction in cost \cite{dudekula2012benefits}.\end{tabular} \\ \hline
    \begin{tabular}[c]{@{}l@{}}$Automated Tests\to$\\ $Bug Detection$\end{tabular}        & \begin{tabular}[c]{@{}l@{}}Automated tests are related to increased fault detection \cite{dudekula2012benefits}.\end{tabular} \\ \hline
    \begin{tabular}[c]{@{}l@{}}$Automated Tests\to$\\ $Technical Challenges$\end{tabular} & \begin{tabular}[c]{@{}l@{}}Automated tests require different skills to implement them\\effectively \cite{garousi2016when}.\end{tabular} \\ \hline
    \begin{tabular}[c]{@{}l@{}}$LackTechKnowledge$\\ $\to Automated Tests$\end{tabular}   & \begin{tabular}[c]{@{}l@{}}The skills level of testers could be a hindrance to test\\automation \cite{garousi2016when}.\end{tabular} \\ \hline
    %\begin{tabular}[c]{@{}l@{}}$Stability\to$\\ $Technical Challenges$\end{tabular}       & \begin{tabular}[c]{@{}l@{}}The stability and maturity of the software under test affect\\the maintenance effort of tests \cite{garousi2016when}.\end{tabular} \\ \hline
    \begin{tabular}[c]{@{}l@{}}$Test Design\to$\\ $Test Reusability$\end{tabular}              & \begin{tabular}[c]{@{}l@{}}Designing tests with maintenance in mind, they can be repeated\\frequently \cite{dudekula2012benefits}.\end{tabular} \\ \hline
    \begin{tabular}[c]{@{}l@{}}$Test Repetition\to$\\ $Reliability$\end{tabular}          & \begin{tabular}[c]{@{}l@{}}When repeating tests, they are more reliable than single\\executions \cite{dudekula2012benefits}.\end{tabular} \\ \hline
    \begin{tabular}[c]{@{}l@{}}$Proj Age\to$\\ $Automated Tests$\end{tabular}             & \begin{tabular}[c]{@{}l@{}}The number, coverage, and maturity of automated tests increase\\with time \cite{zaidman2008mining,zaidman2010studying,hilton2018large,sizilio2019empirical}.\end{tabular} \\ \hline
  \end{tabular}
\end{center}
\end{table}

\begin{figure}[htbp]
  \centerline{\includegraphics[scale=.45]{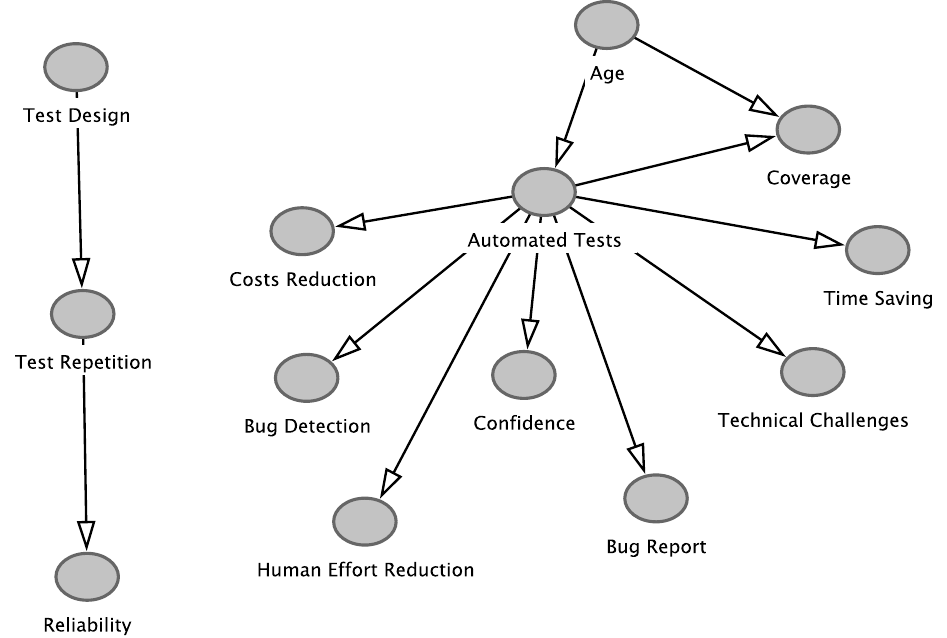}}
  \caption{Partial causal DAG for automated tests associations.}
  \label{fig:dag-tests}
\end{figure}

\begin{figure}[htbp]
  \centerline{\includegraphics[scale=.5]{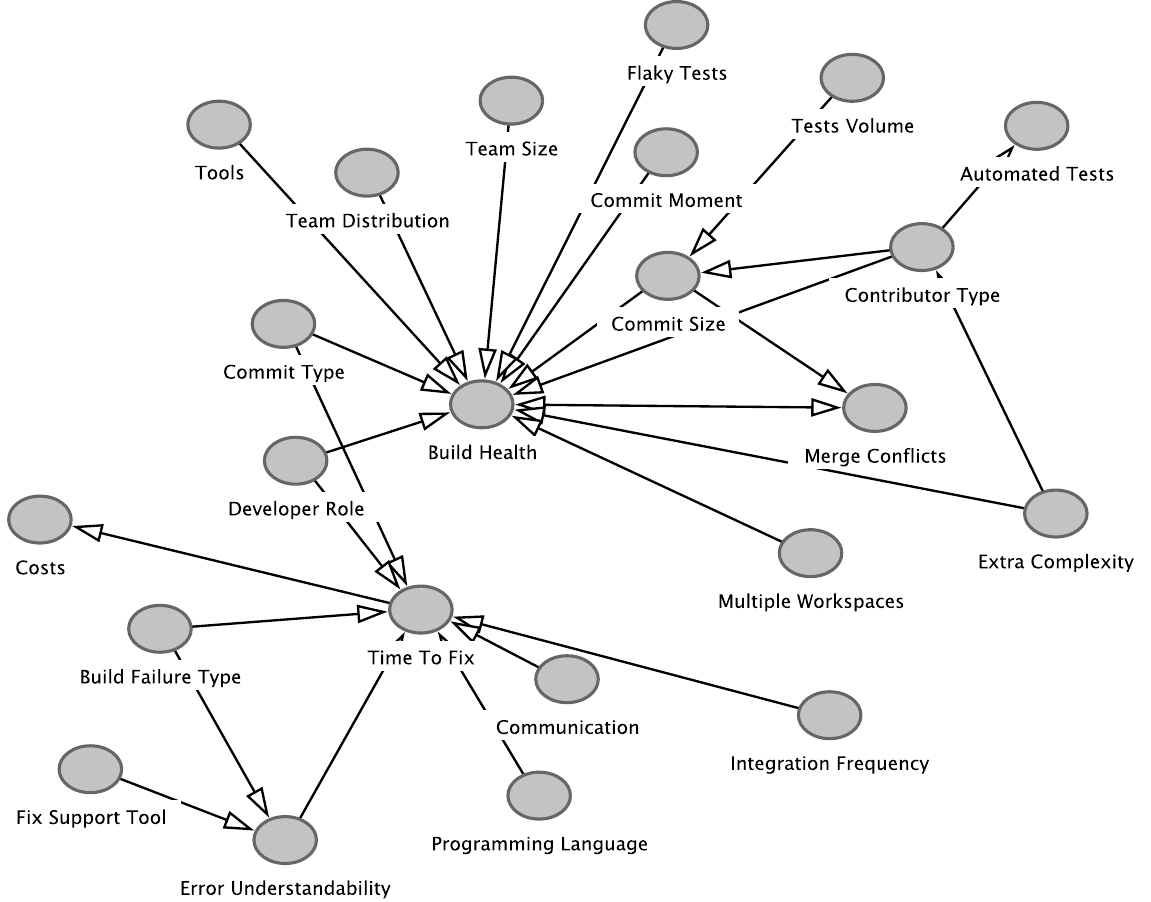}}
  \caption{Partial causal DAG for build attributes and their connections.}
  \label{fig:dag-builds}
\end{figure}

With connections shown in Table~\ref{tab:tests-associations} we build the partial causal DAG represented in Fig.~\ref{fig:dag-tests}, where $Automated Tests$ is the intervention. The relationship $Automated Tests\to Bug Report$ was omitted in this figure because it is a partial DAG centered in the $Automated Tests$. With Table~\ref{tab:build-associations}, we build the partial causal DAG represented in Fig.~\ref{fig:dag-builds}, centered in the build attributes.

\begin{figure*}[htbp]
  \centerline{\includegraphics[scale=.45]{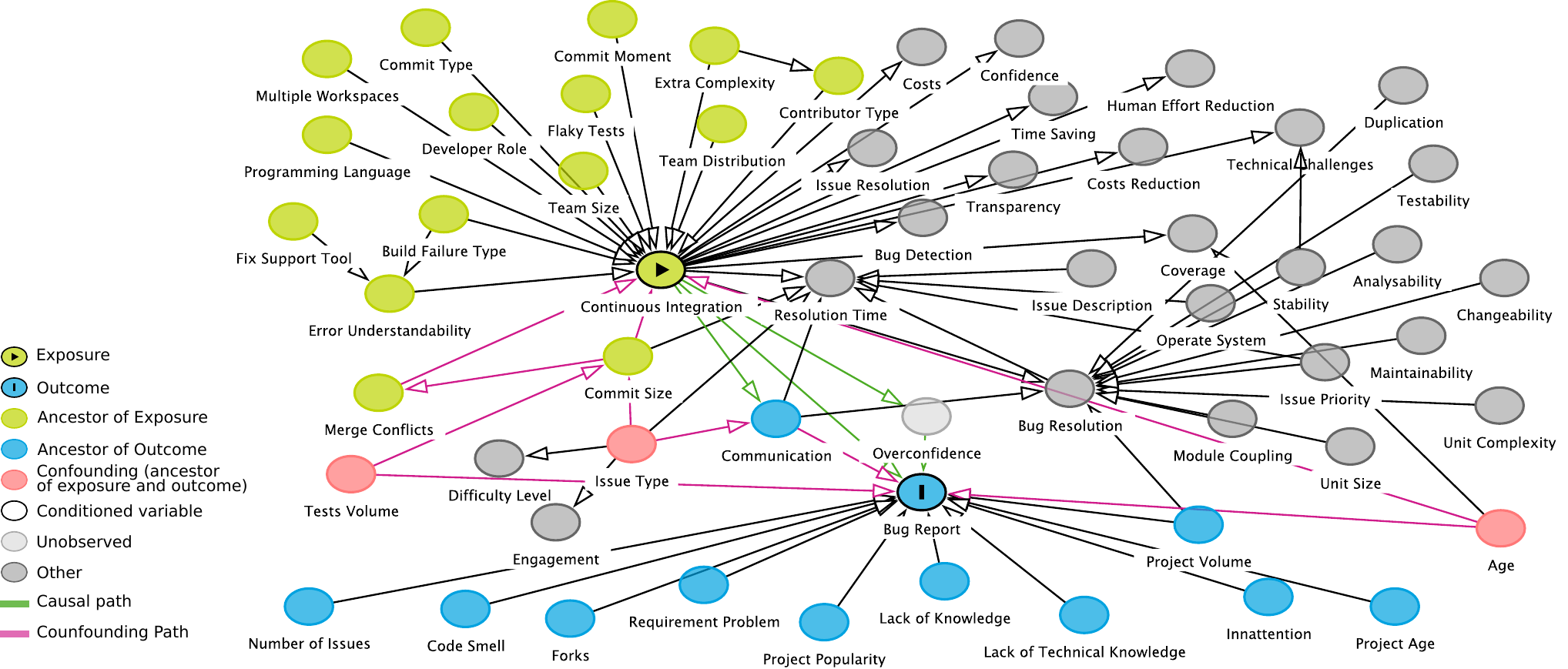}}
  \caption{Complete literature-based causal DAG for CI, Bug Reports and their co-variables.}
  \label{fig:dag-unified}
\end{figure*}

\begin{table}[]
  \caption{CI connections cataloged among the literature from the perspective of build practices.}
  \label{tab:build-associations}
  \begin{center}
  \begin{tabular}{|l|l|}
    \hline
    \textbf{Connections} & \textbf{Rationale}                                                                                                                                                                               \\ \hline
    \begin{tabular}[c]{@{}l@{}}$Build Health\to$\\ $Work Time$\end{tabular}         & \begin{tabular}[c]{@{}l@{}}Broken builds lead to loss of time by freezing development and\\tests \cite{kerzazi2014why}.\end{tabular} \\ \hline
    \begin{tabular}[c]{@{}l@{}}$Build Health\to$\\ $Merge Conflicts$\end{tabular}   & \begin{tabular}[c]{@{}l@{}}Broken builds lead to work blockage, which in turn leads to merge\\conflicts \cite{laukkanen2017problems}.\end{tabular} \\ \hline
    \begin{tabular}[c]{@{}l@{}}$Team Size\to$\\ $Build Health$\end{tabular}         & \begin{tabular}[c]{@{}l@{}}The team size relates to build breakage. Shorter teams tend to break\\fewer than larger ones \cite{kerzazi2014why}.\end{tabular} \\ \hline
    \begin{tabular}[c]{@{}l@{}}$Multiple Workspace$\\ $\to Build Health$\end{tabular}& \begin{tabular}[c]{@{}l@{}}Maintaining multiple physical structures for multiple branches is\\associated with more build breakage \cite{kerzazi2014why}.\end{tabular} \\ \hline
    \begin{tabular}[c]{@{}l@{}}$Developer Role\to$\\ $Build Health$\end{tabular}    & \begin{tabular}[c]{@{}l@{}}There is a statistical difference in build breakage among different\\role groups \cite{kerzazi2014why}.\end{tabular} \\ \hline
    \begin{tabular}[c]{@{}l@{}}$Commit Size\to$\\ $Build Health$\end{tabular}       & \begin{tabular}[c]{@{}l@{}}The size of the changes is related to a higher probability of build\\failure \cite{kerzazi2014why,islam2017insights,rausch2017empirical}.\end{tabular} \\ \hline
    \begin{tabular}[c]{@{}l@{}}$Commit Type\to$\\ $Build Health$\end{tabular}       & \begin{tabular}[c]{@{}l@{}}The commit type (such as features and bugs) and the contribution\\model (e,g., pull request and push model) are associated with build\\breakage \cite{kerzazi2014why,islam2017insights,rausch2017empirical}.\end{tabular} \\ \hline
    \begin{tabular}[c]{@{}l@{}}$Commit Moment\to$\\ $Build Health$\end{tabular}     & \begin{tabular}[c]{@{}l@{}}There is an association between the moment of contributions and the\\rate of build breakage \cite{kerzazi2014why}.\end{tabular} \\ \hline
    \begin{tabular}[c]{@{}l@{}}$Team Distribution$\\ $\to Build Health$\end{tabular} & \begin{tabular}[c]{@{}l@{}}The geographical distance of the team members is associated with\\the build results \cite{kerzazi2014why}.\end{tabular} \\ \hline
    \begin{tabular}[c]{@{}l@{}}$Tools\to$\\ $Build Health$\end{tabular}             & \begin{tabular}[c]{@{}l@{}}The languages and their tools are related to different build\\breakage rates \cite{seo2014programmers}.\end{tabular} \\ \hline
    \begin{tabular}[c]{@{}l@{}}$Extra Complexity$\\ $\to Build Health$\end{tabular}  & \begin{tabular}[c]{@{}l@{}}Complex builds tend to break \cite{laukkanen2017problems}.\end{tabular} \\ \hline
    %\begin{tabular}[c]{@{}l@{}}$Merge Conflicts\to$\\ $Build Health$\end{tabular}   & \begin{tabular}[c]{@{}l@{}} \cite{laukkanen2017problems}.\end{tabular} \\ \hline
    \begin{tabular}[c]{@{}l@{}}$Flaky Tests\to$\\ $Build Health$\end{tabular}       & \begin{tabular}[c]{@{}l@{}}Flaky tests favor the occurrence of build breakage \cite{laukkanen2017problems,rausch2017empirical}.\end{tabular} \\ \hline
    \begin{tabular}[c]{@{}l@{}}$Contributor Type\to$\\ $Build Health$\end{tabular}  & \begin{tabular}[c]{@{}l@{}}Less frequent contributors tend to break builds less \cite{rausch2017empirical}.\end{tabular} \\ \hline
    \begin{tabular}[c]{@{}l@{}}$Time To Fix\to$\\ $Costs$\end{tabular}              & \begin{tabular}[c]{@{}l@{}}The time lost relates directly to a monetary cost \cite{kerzazi2014why}.\end{tabular} \\ \hline
    \begin{tabular}[c]{@{}l@{}}$Communication\to$\\ $Time To Fix$\end{tabular}      & \begin{tabular}[c]{@{}l@{}}The feedback mechanisms and information speed affect the\\awareness of a broken build and the time to fix it \cite{kerzazi2014why}.\end{tabular} \\ \hline
    \begin{tabular}[c]{@{}l@{}}$Developer Role\to$\\ $Time To Fix$\end{tabular}     & \begin{tabular}[c]{@{}l@{}}The developer role is associated with the time to fix a broken\\build \cite{kerzazi2014why}.\end{tabular} \\ \hline
    \begin{tabular}[c]{@{}l@{}}$Commit Type\to$\\ $Time To Fix$\end{tabular}        & \begin{tabular}[c]{@{}l@{}}The characteristics of the branches and code access (e.g., isolated\\branches) are associated with the time to fix a broken build \cite{kerzazi2014why}.\end{tabular} \\ \hline
    \begin{tabular}[c]{@{}l@{}}$Integration Freq$\\ $\to Time To Fix$\end{tabular}& \begin{tabular}[c]{@{}l@{}}The integration frequency in the team affects the build fixing \cite{kerzazi2014why}.\end{tabular} \\ \hline
    \begin{tabular}[c]{@{}l@{}}$Program Language$\\ $\to Time To Fix$\end{tabular}   & \begin{tabular}[c]{@{}l@{}}The programming language is related to the time spent to fix\\a broken build \cite{seo2014programmers}.\end{tabular} \\ \hline
    \begin{tabular}[c]{@{}l@{}}$Error Understand$\\ $\to Time To Fix$\end{tabular}   & \begin{tabular}[c]{@{}l@{}}The understandability of the build failures directly impacts\\the time needed to solve them \cite{vassalo2014every}.\end{tabular} \\ \hline
    \begin{tabular}[c]{@{}l@{}}$Build Fail Type$\\ $\to Time To Fix$\end{tabular}   & \begin{tabular}[c]{@{}l@{}}The build failure types are associated with different difficulty\\levels \cite{vassalo2014every}.\end{tabular} \\ \hline
    \begin{tabular}[c]{@{}l@{}}$Tests Volume\to$\\ $Commit Size$\end{tabular}   & \begin{tabular}[c]{@{}l@{}}Complex and time-consuming testing is a possible reason for large\\commits \cite{laukkanen2017problems}.\end{tabular} \\ \hline
    \begin{tabular}[c]{@{}l@{}}$Contributor Type\to$\\ $Commit Size$\end{tabular}   & \begin{tabular}[c]{@{}l@{}}The type of contributor (e.g., casual) relates to the build\\breakage rate \cite{reboucas2017how}.\end{tabular} \\ \hline
    \begin{tabular}[c]{@{}l@{}}$Contributor Type\to$\\ $Automated Tests$\end{tabular}   & \begin{tabular}[c]{@{}l@{}}The contributor type is related to the number of automated\\tests \cite{reboucas2017how}.\end{tabular} \\ \hline
    \begin{tabular}[c]{@{}l@{}}$Extra complexity\to$\\ $Contributor Type$\end{tabular}   & \begin{tabular}[c]{@{}l@{}}The complexity of the jobs is related to the type of contributor\\in the projects \cite{reboucas2017how}.\end{tabular} \\ \hline
    \begin{tabular}[c]{@{}l@{}}$Commit Size\to$\\ $Merge Conflicts$\end{tabular}   & \begin{tabular}[c]{@{}l@{}}Large commits are associated with merge conflicts \cite{laukkanen2017problems}.\end{tabular} \\ \hline
    \begin{tabular}[c]{@{}l@{}}$Fix Tools\to$\\ $Error Understand$\end{tabular}   & \begin{tabular}[c]{@{}l@{}}Fix support tools improves the understandability of the build\\logs \cite{vassalo2014every}.\end{tabular} \\ \hline
    \begin{tabular}[c]{@{}l@{}}$Build Fail Type\to$\\ $Error Understand$\end{tabular}   & \begin{tabular}[c]{@{}l@{}}The build failure type is associated with different levels of\\understandability \cite{vassalo2014every}.\end{tabular} \\ \hline

  \end{tabular}
\end{center}
\end{table}

Fig.~\ref{fig:dag-unified} shows the unified and complete literature-based causal DAG, i.e., the union of the assumptions cataloged in 
Tables~\ref{tab:ci-bug-associations},~\ref{tab:bugreports-associations},~\ref{tab:bugresolution-associations},~\ref{tab:resolutiontime-associations},~\ref{tab:internal-bugs-associations},~\ref{tab:tests-associations},~\ref{tab:build-associations}.
The literature-based causal DAG expresses all known associations for the related variables in the studied domain. 
The nodes on the DAG represent variables (e.g., Continuous Integration and Bug Report), and the directed edges connecting the variables represent an association between them. The associations ``flow'' from one variable to another. For instance, according to Fig.~\ref{fig:dag-unified}, Continuous Integration influences Bug Report, i.e., CI teams discover more bugs and CI projects present fewer defects than NOCI projects (those that do not adopt CI)~\cite{amrit2018effectiveness,vasilescu2015quality}. Such DAG associations could represent a positive or a negative association, and whenever possible, this interpretation is discussed in the text.
Note that the variables $Automated Tests$, $Build Health$, $Integration Frequency$, and $Time To Fix$ are all represented by $Continuous Integration$. This causal DAG built upon the existing literature is the starting point for identifying variables that must be measured and controlled to allow a causal analysis of the influence of CI on software quality.
\\

\subsubsection{\textbf{DAG Building}}
%We built the literature-based DAG in Fig.~\ref{fig:dag-unified} expressing the associations from Tables~\ref{tab:ci-bug-associations},~\ref{tab:bugreports-associations},~\ref{tab:bugresolution-associations},~\ref{tab:resolutiontime-associations},~\ref{tab:internal-bugs-associations},~\ref{tab:tests-associations}, and~\ref{tab:build-associations}.
To analyze the causal effect of CI on software quality, we need to be attentive to all potential confounding effects, i.e., bias due to backdoor paths (see Section~\ref{confounding}). Thus, we draw a DAG containing a sufficient set of variables that show the existing backdoor paths with respect to CI and software quality. This new DAG is a subgraph of the DAG shown in Fig.~\ref{fig:dag-unified}.

Based on Reichenbach's Common Cause Principle and the Markov Condition, we know that a causal DAG should include the common causes of any pair of variables in the DAG \cite{pearl2000models,hernan2010}. Therefore, we can discard several variables in Fig.~\ref{fig:dag-unified} because they are external variables without connecting with the variables under investigation (i.e., $CI$ and $Bug Report$), and they are not common causes for any variable selected for the analysis (i.e., CI, Bug Report or one of its common causes). For instance, $Project Popularity$  is associated with $Bug Report$ but does not have an association with no other variable on the DAG. Thus, $Project Popularity$ can be discarded from our causal analysis. On the other hand, $Age$ is a common cause associated with $Continuous Integration$ and $Bug Report$, and it is essential to analyze the causal effect between them since $Age$ is a potential confounding factor (see Section~\ref{confounding}).

\subsection{\textbf{RQ2. \RQtwo} \label{subrq2}}
Based on the concepts of conditional independence and the d-separation rules (see Section~\ref{d-separation}), we analyze a set of statistically testable restrictions as implications of a model \cite{pearl2000models}. Such restrictions (i.e., statistical implications) are conditional or unconditional independencies between DAG variables that must be found in any dataset generated by the causal processes described in the DAG. By building a dataset with the variables in our causal DAG, we can test the implications of the DAG statistically. Note that because these statistical tests are based on the d-Separation rules, we are not only checking for associations but also for causal relationships (i.e., considering the ``flow'' between relationships in the DAG)~\cite{pearl1991theory,pearl1995causal,pearl2000models}. We are discovering the causal structure and inferring causation from the initial DAG built in the RQ1 (section~\ref{subrq1}) and the testing of their d-Separation implications on an empirical data set~\cite{shalizi2021}.

The implications are in the form of unconditional independencies, like $Age \CI Tests Volume$, which means $Age$ is independent of $Tests Volume$. Alternatively, the implications may have the form of conditional independencies, like $Merge Conflicts \CI Tests Volume \mid Commit Frequency$, which means $Merge Conflicts$ is independent of $Tests Volume$ conditioned in $Commit Frequency$. Since we use the \texttt{DAGitty R} package~\cite{textor2017dagitty} to draw our causal DAGs, we obtain from the \texttt{impliedConditionalIndependence} function a list of testable implications that become our causal hypotheses.

We then collect data by mining software repositories to empirically analyze the selected variables, allowing us to test the causal DAG and its statistical implications, i.e., our causal hypotheses.

\begin{comment}
\begin{itemize}
	\item H1. $Age\CI Commit Size$
	\item H2. $Age\CI Merge Conflicts$
	\item H3. $Age \CI Tests Volume$
	\item H4. $Age \CI Communication  \mid$ $Commit Size$, $Continuous Integration$
	\item H5. $Merge Conflicts \CI Communication  \mid$ $Commit Size$, $Continuous Integration$
	\item H6. $Merge Conflicts \CI Tests Volume  \mid$ $Commit Size$
	\item H7. $Merge Conflicts \CI Bug Report  \mid$ $Age$, $Communication$, $Continuous Integration$, $Tests Volume$
	\item H8. $Merge Conflicts\CI Bug Report \mid$ $Age$, $Commit Size$, $Continuous Integration$
	\item H9. $Continuous Integration \CI Tests Volume \mid$ $Commit Size$
	\item H10. $Bug Report \CI Commit Size \mid$ $Age$, $Communication$, $Continuous Integration$, $Tests Volume$	
\end{itemize}
\end{comment}

\subsubsection{\textbf{Collecting Data \& Empirical Analysis \label{msr}}}

\begin{figure*}[htbp]
  \centerline{\includegraphics[scale=.45]{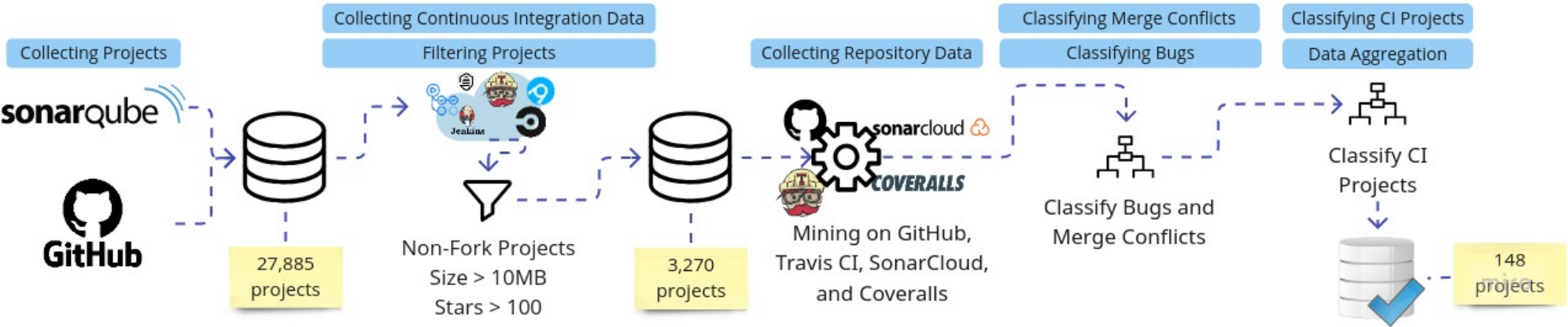}}
  \caption{Mining Software Repository Process}
  \label{fig:msr}
\end{figure*}

Based on the knowledge and assumptions acquired in the literature review and the causal DAG built to answer the RQ1, we have support for building a dataset to analyze the relationship between CI and software quality, including confounding variables. Using the dataset, we can check the validity of our DAG through statistical tests. Fig.~\ref{fig:msr} summarizes the dataset building process, and we detail such process in the following. All tools, scripts, and information necessary to reproduce the process described in the sequence are available in our replication package~\footnote{\replication}.
%Thus we start mining 27885 popular GitHub\footnote{https://github.com} software repositories and investigate if they are using some CI service. We discovered a high CI service usage, being Travis CI\footnote{https://travis-ci.org/} is the most popular. We then select projects that have:  (i) more than 100 stars; (ii) size greater than 10MB; (iii) are not forking ones; (iv) using Travis CI or nor CI service (i.e., discard projects using other services). 

%After applying these criteria, our sample was reduced to 2,285 repositories, 1,515 with Travis CI and 770 with no CI service. For these repositories, we mined 1,425,493 pull requests and 1,050,371
%issues. We screen projects considering only those with 12 consecutive active months with merged pull requests and closed issues. We obtained a new subset with 537 repositories.

%For these 537 projects, we mined the pull request commits, build and coverage information, and classified merge conflicts and bug issues. To classify a project as CI, we consider the coverage and build information beyond the Travis CI usage. We achieve, then, a final dataset containing the variables of interest for 70 projects---35 non-CI and 35 CI projects. In Section~\ref{rq2} we better detail the dataset and answer the RQ2.

\textbf{Collecting Projects. }
As illustrated in Fig.~\ref{fig:msr}, we start with a list of 27,885 projects obtained from the most starred repositories in the GitHub Search API\footnote{https://docs.github.com/en/rest/reference/} and a list of projects' names from the SonarQube web API\footnote{https://community.sonarsource.com/t/list-of-all-public-projects-on-sonarcloud-using-api/33551}. We searched for projects having a public repository on GitHub to start the mining repository process by collecting the history of developers' contributions to them. The search included 15 popular languages: C, C\#, C++, Go, Java, JavaScript, Kotlin, Objective-C, PHP, Python, Ruby, Rust, Scala, Swift, and TypeScript. So the only initial premise was to have a public repository on GitHub.

\textbf{Collecting Continuous Integration Data. }
In the next steps, we search for data in continuous integration services. Inspired by Hilton's work~\cite{hilton2016usage}, which searches for the usage of 5 different services --- TravisCI, CircleCI, AppVeyor, Werker, Cloud-Bees, and Jenkins CI, we search for the same services and add a new one --- GitHub Actions\footnote{https://github.com/features/actions}. We discovered a high CI service usage (54.3\%) among the collected projects. Travis CI~\footnote{https://travis-ci.org/} appeared as the most popular service in our dataset. 

Despite the evidence regarding a decreasing usage of Travis CI service~\cite{widder2018leaving,decan2022,golzadeh2022}, it is worth mentioning that Travis is still the most popular service in our dataset (see Table~\ref{tab:ci-services}), which has also been the case for recent research analyzing machine learning projects using CI~\cite{rzig2022}. As for GitHub Actions, although the relatively short time since its launch (November 2019), it has been used by a significant number of projects up to our study's data collection date (2022). This observation confirms the findings from Decan et al.~\cite{decan2022}. When they analyzed 67,870 GitHub repositories, they found that more than 4 out of 10 use GitHub Actions workflows. Golzadeh et al.~\cite{golzadeh2022} also observed that while the adoption rate of the other CI services decreased, GitHub Actions has a steadily increasing adoption rate.

Table~\ref{tab:ci-services} displays the identified Continuous Integration (CI) services, the number of projects categorized as users for each service, and the criteria employed to categorize and classify the adoption of these services.

We build scripts to identify which CI service is used by each project. For Travis CI, Circle CI, and Wercker, we check the public APIs of these services by searching for the project name and organization. Specifically for Travis CI, if we do not find any information through the Travis API, our scripts search for the \textit{``.travis.yml''}  file using the GitHub API. We also keep track of when the service configuration was initiated, specifically noting the date of the first commit of the \textit{``.travis.yml''} file.

For the remaining services, the script checks if the configuration file exists. In the case of AppVeyor, we verify the presence of a file named \textit{``appveyor.yml''} within the GitHub repository using the API. For Jenkins, we search for a \textit{``Jenkinsfile''}, while for Circle CI, we look for a \textit{``config.yml''} inside a \textit{``.circleci''} folder. Similarly, for GitHub Actions, we examine whether a file with the extension \textit{``.yml''} exists inside a \textit{``.github/workflows''} folder.

\begin{table}[H]
	\caption{The CI service usage on the dataset and the classification criteria.}
	\label{tab:ci-services}
	\begin{tabular}{|l|l|l|}
	\hline
	\textbf{CI Service}                                       & \textbf{Qty} & \textbf{Criteria}                                                                                                                                                                                                                                                                                                               \\ \hline
	Travis CI                                                 & \begin{tabular}[c]{@{}l@{}}9,092\\(32.6\%)	\end{tabular}       & \begin{tabular}[c]{@{}l@{}}We search on Travis API$^{\mathrm{a}}$ for the existence of the project in the service.\\If true, we search on GitHub for the existence of a \textit{``.travis.yml''} file.\end{tabular}                    \\ \hline
	\begin{tabular}[c]{@{}l@{}}GitHub \\ Actions\end{tabular} & \begin{tabular}[c]{@{}l@{}}5,630\\(20.1\%)	\end{tabular}       & \begin{tabular}[c]{@{}l@{}}Using the search code feature of GitHub API, we search for the\\extension \textit{``.yml''} in the path \textit{``.github/workflows''}.\end{tabular}                                                                                                             \\ \hline
	Circle CI                                                 & \begin{tabular}[c]{@{}l@{}}307\\(1.1\%)		\end{tabular}       & \begin{tabular}[c]{@{}l@{}}Through Circle CI API$^{\mathrm{b}}$ we search for the existence of the project on\\the service. If not located, we searched for the file \textit{``config.yml''} in the\\path \textit{``.circleci''}.\end{tabular} \\ \hline	
	Jenkins                                                   & \begin{tabular}[c]{@{}l@{}}58\\(0.2\%)		\end{tabular}       & \begin{tabular}[c]{@{}l@{}}Using the search code feature of GitHub API, we search for the file\\ \textit{``Jenkinsfile''}.             \end{tabular}                                                                                                                                                                                                     \\ \hline
	AppVeyor                                                  & \begin{tabular}[c]{@{}l@{}}29\\(0.1\%)		\end{tabular}       & \begin{tabular}[c]{@{}l@{}}Using the search code feature of GitHub API, we search for the file\\ \textit{``appveyor.yml''}.         \end{tabular}                                                                                                                                                                                                        \\ \hline
	Wercker                                                   & 0            													& \begin{tabular}[c]{@{}l@{}}Using the Wercker API$^{\mathrm{c}}$ we search for the existence of the project on\\the service.\end{tabular}                                                                                                                 \\ \hline
	\begin{tabular}[c]{@{}l@{}}No CI\\Service\end{tabular}                                                         & \begin{tabular}[c]{@{}l@{}}12,769\\(45.7\%)	\end{tabular}       &                                                                                                                                                                                                                                                                                                                                 \\ \hline
	\multicolumn{3}{l}{$^{\mathrm{a}}$https://docs.travis-ci.com/user/developer/\#api-v3}\\
	\multicolumn{3}{l}{$^{\mathrm{b}}$https://circleci.com/docs/api/v2/}\\
	\multicolumn{3}{l}{$^{\mathrm{c}}$https://devcenter.wercker.com/development/api/endpoints/}\\
	\end{tabular}
\end{table}

\textbf{Filtering projects. }
%Afterwards, we filtered out irrelevant repositories as represented in Fig.~\ref{fig:msr}. In particular, we consider only non-fork projects with more than 100 stars. To select ``engineered projects''~\cite{munaiah2017curating} as opposed to toy projects, we apply a series of criteria, beginning by excluding projects smaller than 10MB~\cite{oliveira2017draco,oliveira2019finding}. These two criteria are the initial steps towards selecting projects that approach the ``engineered projects'' concept~\cite{munaiah2017curating}. While mining data from the repository and classifying projects for analysis, we ensure a rigorous selection of projects that represent engineered projects. In the sequence, we detail each step of mining and classification of projects.
Afterwards, we filtered out irrelevant repositories as represented in Fig.~\ref{fig:msr}. In particular, we consider only non-fork projects with more than 100 stars. Despite the risks of applying MSR in empirical studies and selecting non-representative projects~\cite{munaiah2017curating}, we selected ``engineered projects'' as opposed to toy projects~\cite{munaiah2017curating} and applied a series of criteria, beginning by excluding projects smaller than 10MB~\cite{oliveira2017draco,oliveira2019finding}. These two criteria are the initial steps towards selecting projects that approach the ``engineered projects'' concept~\cite{munaiah2017curating}. While mining data from the repository and classifying projects for analysis, we ensure a rigorous selection of projects that represent engineered projects. In the sequence, we detail each step of mining and classification of projects.

Next, we consider only projects using either Travis CI or projects not using a CI service. Including projects not using a CI service is essential because we create a control group of no-CI projects in order to understand the effects on the CI projects group. As mentioned earlier, we selected Travis CI as a target because it is the most used service in our dataset. In addition, Travis CI has a public API to obtain software builds data. After this filtering step, our sample was reduced from 27,885 to 3,270 repositories, i.e., 2,527 repositories using Travis CI and 743 not using a CI service.
 
\textbf{Collecting Repository Data. } 
We mine the pull requests and issues of the projects, filtering out those projects that do not have pull requests or issues. We collect data from pull requests, issues, commits, and comments using the GitHub Search API. We use the Travis API to collect data from software builds, while we use two different services (Coveralls~\footnote{https://coveralls.io/} and SonarCloud~\footnote{https://sonarcloud.io/}) to search for code coverage information.

We collected pull requests and issues in the repositories using the endpoint ``pulls'' and ``issues'' from the GitHub Search API. This process resulted in a total of 1,425,493 pull requests and 3,328,221 issues. 
Next, we collected all pull request comments and associated information regarding authors. All these data allow us to compute a metric to measure $Communication$ detailed forward.

We also collect each pull request's commit, sha, date, message, and author. To collect the detailed commit information regarding lines of code added, removed, or changed, we request it to GitHub API~\footnote{https://api.github.com/repos/\{owner\}/\{repo\}/pulls/\{pull\_number\}/files}. We processed the data to obtain the size of a commit, the number of lines of added/removed code, the number of files modified, and the test files and test lines. The commits data help us to compute metrics to measure $Commit Frequency$,  $Tests Volume$, and $Merge Conflicts$. 

\textbf{Classifying Merge Conflicts.}
To identify the merge conflicts occurrences, we used an algorithm (see Algorithm~\ref{alg:merge_conflicts}) to reconstruct the commits history and identify a merge conflict when it was generated. The algorithm clones the repository (line 1) and gets the list of commits (line 4). The algorithm gets the parents for each commit and verifies the number of parent commits (line 6). If a commit has more than one parent, we call the \texttt{git diff} function to them (line 7). Then, depending on the \texttt{diff} result, we can detect a conflict (line 9) and append it to the list of merge conflicts (line 10). We manually validated the algorithm with a random sample of 40 commits and achieved an accuracy of 90\%.

  \begin{algorithm}
\caption{Detect Merge Conflicts}
\label{alg:merge_conflicts}
\begin{algorithmic}[1]
  \REQUIRE $remote$: URL for the project repository
  \REQUIRE $repository\_path$: Local path to clone the project repository

\STATE $\text{repo} \gets Repo.clone\_from(remote, repository\_path)$
\STATE $merge\_conflicts \gets \emptyset$

\STATE \quad \FORALL{$commit$ in $repo$.$commits$()}

  \STATE  \IF{$commit.parents > 1$}
    
    \STATE $merge\_diff \gets repo.git.diff(commit.parents[0],commit.parents[1])$

    \STATE \IF{$merge\_diff$ \texttt{contains} $"<<<<<<< HEAD"$ and $"<<<<<<<"$}

      \STATE Append $commit.sha$ to $merge\_conflicts$
    \STATE \ENDIF
  \STATE \quad \quad \ENDIF
\STATE \quad \ENDFOR
\STATE \quad \RETURN $merge\_conflicts$
\STATE End function
\end{algorithmic}
\end{algorithm}

\textbf{Classifying Bugs.}
To classify issues as bugs, we consider only projects using GitHub issues and labeling them. We mapped the labels used in their repositories through a semi-manual inspection, i.e., we began with a script to a preliminary parse relying on a list of bug-related keywords~\cite{vasilescu2015quality,jadson2022}. Then, we manually validated the labels for each repository using GitHub issues that have assigned to them. This approach is similar to the approach used by Vasilescu et al.~\cite{vasilescu2015quality} and Santos~\cite{jadson2022}.

The exception to our primary approach is a set of 10 projects with many issues without labels. To avoid missing these data, we adopt a conservative approach classifying as bugs issues containing the terms ``bug'' or ``fix'' in their title or their body. %We classified 90,159 out of 3,328,221 issues as bugs, representing 8.5\% of occurrences.

\textbf{Classifying CI Projects.}
In order to avoid the CI Theater~\cite{felidre2019citeather,fowler_citeather} and also to adopt recommendations from Soares et al. ~\cite{soares2021effects}, in addition to the use of Travis CI, we consider build and code coverage information, meaning that we select projects that have actual build and test activity.
After achieving a set of 74 CI projects surviving all filtering stages, we randomly drew 74 out of 95 no-CI projects to balance our dataset. Table~\ref{tab:dataset} shows a summary of the final dataset.

\textbf{Data Aggregation.}
With the collected data, we can analyze dimensions of releases such as Commit Frequency, Test Volume, Merge Conflicts, Communication, Bug Reports, and Age variables. We cannot measure Issue Type and Overconfidence for two reasons. First, classifying $Issue Type$ from the collected issue data is challenging given that projects do not have consistent patterns for classifying issue patterns (e.g., standardized tags, such as ``enhancement'', ``perfective'', ``corrective'' used in a standardized manner). Second, overconfidence is a subjective feeling not feasible to measure through collecting data from repositories. Therefore, we consider these two variables as \textit{latent} (i.e., unmeasured). Latent variables can still be represented and analyzed in a causal DAG, even if they cannot be statistically tested. However, we can still interpret these latent variables based on the statistical tests of the other variables within the DAG~\cite{hernan2010}.

In the DAG and the dataset, we used the variable Commit Frequency, replacing Commit Size. These variables have an intrinsic relationship since more frequent commits tend to be smaller. However, it is noteworthy that commit frequency is most noticeable in the context of projects using Travis CI~\cite{zhao2017impact}. In addition, more frequent commits engender a cascade of CI builds and verifications, potentially detecting bugs earlier, as cited in~\cite{soares2021effects}. In light of this, we employed a commit frequency metric to build a more consistent and complete DAG.

We derive the metrics by aggregating their values by project releases. We collected data covering 12 development months, similar to data analyzed by Zhao et al.~\cite{zhao2017impact} and Santos et al.\cite{jadson2022}. Notably, all 148 projects under scrutiny encompass a data-mined span of 12 months. We consider the first 12 months for CI projects starting from the month when CI was adopted. For no-CI projects, we consider the starting month for analysis according to the median $Age$ of the CI projects, similar to Sizilio et al.~\cite{sizilio2019empirical}.

In addition to $Age$, the $Size$ of the projects are also balanced. We tested this difference through a \texttt{Wilcoxon Rank} test and found no statistically significant difference between the two groups ($p-value = 0.7372$). Regarding the difference in bug reports, we also obtained a negligible difference ($p-value = 0.1318$). On the other hand, we found a difference between the two groups for the number of stars on GitHub ($p-value = 0.000679$), for the number of issues ($p-value = 0.007304$), and for pull-requests ($p-value = 1.737{e}{-07}$). In simple terms, the CI projects are more active and popular, but not necessarily with greater maturity and size.

For each project release in such time, we compute the following metrics:
\begin{itemize}
  \item \textbf{Commit Frequency} represents the aggregate count of commits across all pull requests within a given release.
  \item \textbf{Communication} is the sum of comments and review comments in the pull requests. We consider the average of communication in a release.
  \item \textbf{Merge Conflicts} is the number of merge conﬂicts in a release. We traverse the repositories commit tree and use the GIT command diff to verify the merge commits and the conflicts.
  \item \textbf{Age} is the number of days of a repository from its creation until the release date.
  \item \textbf{Test Volume} is the proportion of lines of test code modiﬁed (added, removed, or changed) in the commits. We applied the strategy from Nery et al.~\cite{sizilio2019empirical} to identify test files. We consider the median test volume in a release.
  \item \textbf{Bug Reports} is the number of bugs reported in a release.
  \item \textbf{CI} is a binary categorical variable indicating whether the project uses CI.
\end{itemize}

\begin{table}[H]
	\caption{Summary of the Data Set.}
	\label{tab:dataset}
  \begin{tabular}{lr}
  \hline
  \textbf{Number of Projects}        & 148                         \\ \hline
  \textbf{Number of Pull Requests}   & \multicolumn{1}{l}{18,961} \\ \hline
  \textbf{Number of Commits}         & 59,034                     \\ \hline
  \textbf{Number of Builds}          & 16,619                      \\ \hline
  \textbf{Number of Issues}          & 41,866                      \\ \hline
  \textbf{Number of Bugs}            & 3,199                      \\ \hline
  \textbf{Number of Merge Conflicts} & 403                        \\ \hline
  \end{tabular}
\end{table}

\subsubsection{\textbf{DAG Implications Testing\label{implications}}}

Having the set of testable causal hypotheses obtained in RQ1 and the produced dataset (Section~\ref{msr}), we can investigate if the implications from the causal DAG are actually held in the empirical dataset. We test the unconditional independence causal hypotheses (e.g., $X \CI Y$) using the \texttt{dcov.test} function from the \texttt{energy R} package~\footnote{https://cran.r-project.org/package=energy}. This test verifies if two variables on the dataset are independent. We test the conditional independence causal hypotheses (e.g., $X \CI Y \mid Z$) with the \textit{Kernel conditional independence test} from the \texttt{CondIndTests R} package~\cite{heize2018kci}.  Both tests are non-parametric and suitable for our data.  
%We consider a confidence interval of 95\% and present the results in the second column of Table~\ref{tab:hypothesis-testing}. 

\subsection{\textbf{RQ3.\RQthree}}
  
Considering the investigations in RQ1 and RQ2, we have information regarding which independence tests have passed or failed in our analyses. Such information allows us to propose adaptations to the causal DAG built in RQ1. For example, by testing the data, we can observe a dependency between $Age$ and $Communication$, so we propose a new edge $Age\to Communication$. We build a new causal DAG by combining the knowledge from the literature (RQ1) and the statistical tests on the dataset (RQ2). We call this hybrid approach data-validated causal DAG. The data-validated causal DAG is compatible with the dataset in terms of the independency relationships between the variables. Then, we perform a data-validated approach for causal discovery.

There are dozens of algorithms for causal discovery from data, such as PC~\cite{spirtes2000causation}, that rely on conditional independence tests to discover the causal structure from the data. On the other hand, Cartwight~\cite{cartwright1994causes} argues that causal investigation requires background information too (i.e., causal assumptions)---``No causes in, no causes out''. 

Thus, to implement the data-validated causal DAG, we start getting background information from the literature review and building the initial causal DAG in RQ1 (i.e., causal assumptions from the literature). Next, we combine the knowledge acquired in RQ2 (i.e., the independence relationships that were not present in the data), and finally, we structure our RQ3 procedure in three steps.

\textbf{Step 1.} By analyzing the rejected hypotheses in RQ2 (Section~\ref{rq2}), in conjunction with d-separation rules (Section~\ref{d-separation}), we can infer connections that should be added or removed in the data-validated causal DAG.

\textbf{Step 2.} We generate the data-validated causal DAG after identifying the disconnected vertices and the necessary changes from Step 1. 

\textbf{Step 3.} We test the d-separation implications on the data-validated causal DAG. If any test fails, we return to Step 1 and refine the DAG.

\begin{figure}[htbp]
  \centerline{\includegraphics[scale=.6]{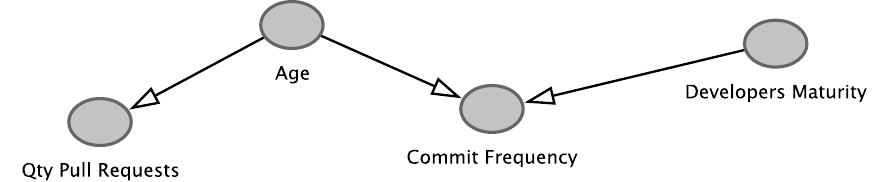}}
  \caption{Hypothetical causal DAG.}
  \label{fig:dag-example-method}
\end{figure}

Fig.~\ref{fig:dag-example-method} presents a hypothetical DAG to illustrate the steps to obtain the data-validated DAG. The DAG contains a Fork structure (i.e., $Qty Pull Requests \leftarrow Age \to Commit Frequency$) and a collider (i.e., $Age \to Commit Frequency \leftarrow DevelopersMaturity$.

\begin{figure}[htbp]
  \centerline{\includegraphics[scale=.6]{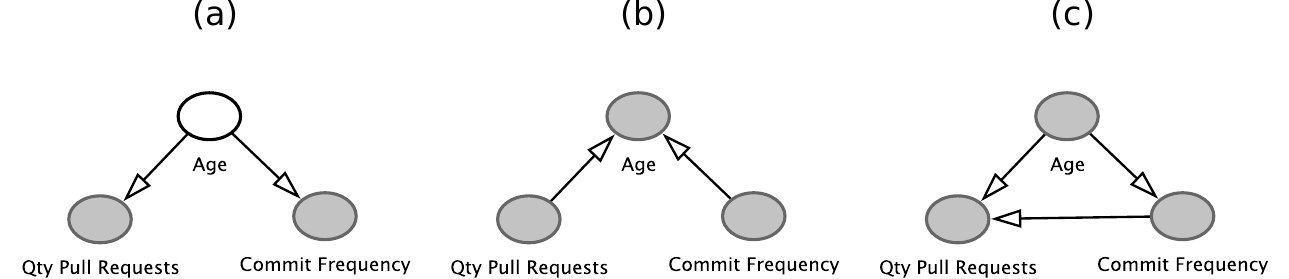}}
  \caption{Examples of causal DAG structure testing.}
  \label{fig:dag-example-method1}
\end{figure}

\textbf{Analyzing Forks and Chains.} The d-Separation rules impose the same principle for the analysis of forks or chains (see Section~\ref{d-separation}), i.e.,  if this structure is true, $Qty Pull Requests$ and $Commit Frequency$ should be statistically dependent because an association flows through forks and chains. Otherwise, these variables should be independent when we block the path by conditioning on the middle variable $Age$, as in Fig.~\ref{fig:dag-example-method1}(a).

Therefore, we can statistically test the fork in Fig.~\ref{fig:dag-example-method1}. We should perform a conditional independence test to check if $Qty Pull Requests$ and $Commit Frequency$ are independent conditioning on $Age$ ($Qty Pull Requests \CI Commit Frequency \mid Age$) as in Fig.~\ref{fig:dag-example-method1}(a). If the test rejects such a hypothesis, meaning that the structure is wrong, we may hypothesize different structures. 

For instance, ``what if the structure would be a collider $Qty Pull Requests \to Age \leftarrow Commit Frequency$? (see Fig.~\ref{fig:dag-example-method1}(b)). In this case, $Qty Pull Requests$ and $Commit Frequency$ would be independent without conditioning. On the other hand, if the variables have a dependence that does not interrupt when conditioning on $Age$, they share a relationship passing through another path, then we add a new edge between $Qty Pull Requests \leftarrow Commit Frequency$, as illustrated in Fig.~\ref{fig:dag-example-method1}(c).

\begin{figure}[htbp]
  \centerline{\includegraphics[scale=.6]{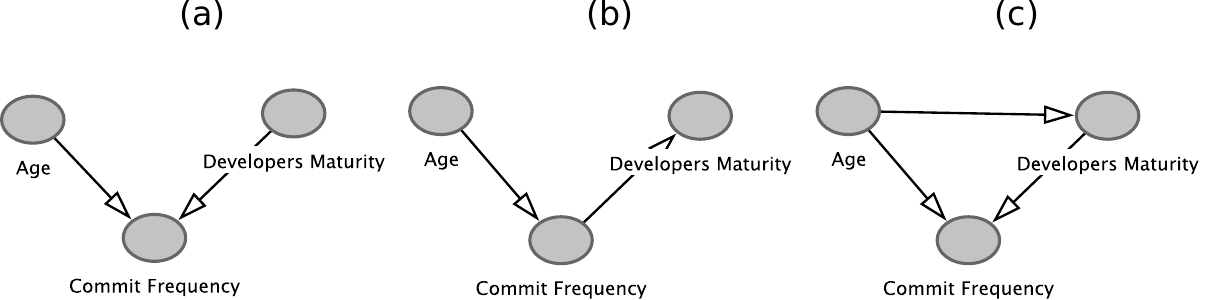}}
  \caption{Examples of causal DAG collider structure testing.}
  \label{fig:dag-example-method2}
\end{figure}

\textbf{Analyzing Colliders.} The d-Separation rules (see Section~\ref{d-separation}) imply that the variables in a collider are independent because the collider blocks the association flow, i.e., $Age$ and $Developers Maturity$ should be statistically independent. On the other hand, if we condition on $Commit Frequency$, they would be dependent because the condition opens the association flow in a collider.

We can statistically test the structure in Fig.~\ref{fig:dag-example-method2}(a) by testing the independence between $Age$ and $Developers Maturity$ ($Age \CI Developers Maturity$). In case the variables are not independent, it means that the structure is not a collider or there exists another path linking them. Thus, we can verify if the structure is a chain as illustrated in Fig.~\ref{fig:dag-example-method2}(b) by verifying if conditioning on $Commit Frequency$ they become independent. Finally, if the last hypothesis fails, i.e., $Age$ and $Developers Maturity$ are not independent, even if conditioning on $Commit Frequency$, then this suggests another path exists between them. In this case, we would add a new edge $Age \to Developers Maturity$, as shown in Fig.~\ref{fig:dag-example-method2}(c).

\section{Results \label{results}}

\subsection{RQ1. \RQone\label{rq1}}

%\subsubsection{Motivation}
%Although there is a belief that Continuous Integration causes an improvement in the software quality and a body of evidence presenting an association, to the best of our knowledge, there is no study assessing the relationship between CI and software quality in a causal perspective~\cite{soares2021effects}. This research question explores the domain literature to catalog the existing assumptions and build a causal DAG containing all relevant literature-reported associations.

%\subsubsection{Approach}
%As explained in Section~\ref{section-literature-survey}, we consider relationships with ``issues'', ``bugs'', or ``defects'' as proxies to software quality. We perform a literature review in an inductive approach for each new variable discovered and inserted in the DAG. Thus, we can know the existing associations between CI and software quality and adjacent associations. This domain knowledge enables us to build a comprehensive DAG having parent variables, descendants, and common causes. Then we can analyze a possible causal relationship between CI and software quality.

%\subsubsection{Results}
Considering the Reichenbach's Common Cause Principle~\cite{reichenbach1991direction} and the Markov Condition \cite{pearl2000models,hernan2010}, in order to produce a sufficient causal DAG, we can exclude any variable that is not a common cause between CI and software quality or one of their ancestors. Thus, we remove from the DAG all non-common cause variables from the causal DAG in Fig.~\ref{fig:dag-unified}, for instance, $Resolution Time$, and $Programming Language$. In addition, we unified the variables $Bug Resolution$ and $Bug Report$, as all connections and the meaning of $Bug Resolution$ are contained in $Bug Report$.

\begin{figure}[htbp]
  \centerline{\includegraphics[scale=.8]{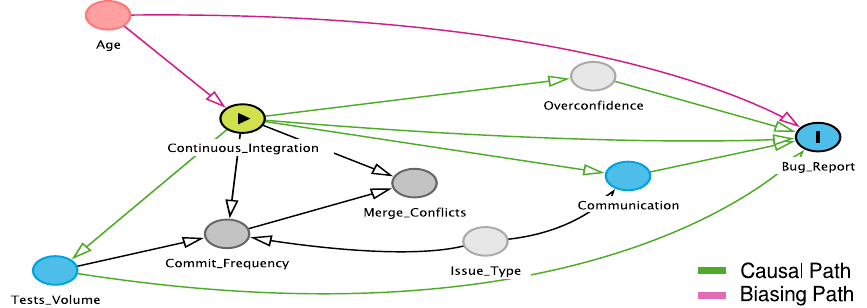}}
  \caption{Final literature-based DAG for CI, Bug Reports and their co-variables.}
  \label{fig:dag-rq1}
\end{figure}

Fig.~\ref{fig:dag-rq1} shows the literature-based DAG after the variable-selection process. According to the DAG, Continuous Integration (CI) has a direct influence on Bug Report~\cite{amrit2018effectiveness,vasilescu2015quality}, as well as an indirect influence through the Communication~\cite{cassee2020silent,huang2012taxonomy} and developers' overconfidence variables~\cite{pinto2017inadequate,pinto2018work,huang2012taxonomy}. That means that CI influences the communication between contributors, which, in turn, communication influences the number of bug reports. A similar phenomenon occurs with developers' overconfidence, but overconfidence tends to have a negative effect on the number of bug reports, i.e., overconfidence may actually increase the number of bug reports or, at least, prevent a team from the opportunity of reducing the number of bug reports. Fig.~\ref{fig:dag-rq1} shows the literature-based DAG after the variable-selection process. According to the DAG, Continuous Integration (CI) has a direct influence on Bug Report~\cite{amrit2018effectiveness,vasilescu2015quality}, as well as an indirect influence through the Communication~\cite{cassee2020silent,huang2012taxonomy} and developers' Overconfidence variables~\cite{pinto2017inadequate,pinto2018work,huang2012taxonomy}. 

The age of a project influences the CI practices~\cite{zaidman2008mining,zaidman2010studying,hilton2018large,sizilio2019empirical} and also influences bug reports~\cite{vasilescu2015quality}, i.e., Age is a common cause for both CI and bug reports, thus opening a backdoor path for bias (see Section~\ref{confounding}). Similarly,  the commit frequency opens other backdoor paths: (i) $Continuous Integration \leftarrow Commit Frequency \leftarrow Tests Volume \to Bug Report$; and (ii) $Continuous Integration \leftarrow Commit Frequency \leftarrow Issue Type \to Communication \to Bug Report$. These backdoor paths represent risks of confounding the influence of CI on other variables such as Bug Report. Thus, it is necessary to condition for some variables blocking these backdoor paths to obtain causal estimations regarding the total causal effect of CI on Bug Report. The minimal conditioning set is formed by $Age$ plus $Commit Frequency$, capable of blocking all backdoor paths (i.e., confounding effects). 

Conditioning for the confounding allows us to measure whether CI has a certain level of causal effect on Bug Reports despite the presence of confounding variables $Age$ or $Commit Frequency$.
% since it is associated with CI practices~\cite{kerzazi2014why,islam2017insights,rausch2017empirical}, with tests volume~\cite{laukkanen2017problems} and with $Issue Type\to Communication$~\cite{hindle2008large,licorish2017exploring}. 

%As we have seen, the ``association flows'' from CI to bug report through other variables. Accordingly to this model, the variables $Commit Size$, $Tests Volume$, and $Age$ are confounding factors providing backdoor paths for associations between $Continuous Integration$ and $Bug Report$, i.e., the influence of CI on $Bug Report$ is confounding by the influence of such variables. Thus, to evaluate a causal influence of CI in the $Bug Report$, we need to adjust for these variables. The minimal adjustment set is formed by $Commit Size$, $Age$, and $Merge Conflicts$, capable of blocking all backdoor paths (i.e., confounding effects) and estimating the total influence of CI in $Bug Report$. Additionally, we can perceive through the DAG that $Communication$ and $Overconfidence$ also mediate this influence.

%\begin{tcolorbox}
%RQ1: The theory-driven causal model shows that in addition to the direct influence of CI on Bug Report, the effect of CI on the communication (among contributors) and developers' overconfidence also affects Bug Report. Thus, the total effect of CI is more extensive than its direct influence. Finally, the CI effect could confound by the influence of the variables tests volume and the project age.
%\end{tcolorbox}

\subsection{RQ2. \RQtwo \label{rq2}}

With the causal DAG obtained in RQ1 (Section~\ref{rq1}) and the dataset we produced (Section~\ref{msr}), we can investigate if the data supports the relationships depicted in the causal DAG. Such evaluation is possible because we can derive statistically testable implications (see Section~\ref{d-separation}) using the d-separation properties on the causal DAG (see Fig.~\ref{fig:dag-rq1}), and we can verify whether the same statistical independency also exists between the variables of the dataset. Thus, we answer RQ2 by testing the dataset against the proposed set of causal hypotheses from Table~\ref{tab:hypothesis-testing-conditional}. To test such causal hypotheses, we use \textit{dcov} test and \textit{Kernel conditional independence test}. 

When necessary, we test the unconditional independence hypotheses (i.e., two variables are independent without conditioning on other variables) using the \textit{dcov} test. This test evaluates if two variables are independent, returning an $R$ value between 0 and 1 when they are independent. If it returns a value greater than 1 the variables are considered to be dependent. For example, the \textit{dcov} test for a hypothesis $H_{n}$ ``$Age\CI Commit Frequency$'' returns an $R$ value of 49.5566 (i.e., larger than 1), meaning that we reject the hypothesis that $Age$ and $Commit Frequency$ are independent.
%We test the unconditional independence hypotheses (i.e., two variables are independent without conditioning on other variables) in Table~\ref{tab:hypothesis-testing-unconditional} using \textit{dcov} test. This test evaluates if two variables are independent, returning an $R$ value between 0 and 1 when they are independent. If it returns a value greater than 1 we assume them as dependents. 

On the other hand, we test the conditional independence hypotheses (i.e., two variables are independent when conditioning on another set of variables) in Table~\ref{tab:hypothesis-testing-conditional} using \textit{Kernel conditional independence test}~\cite{heize2018kci}. This test evaluates the null hypothesis that two variables, Y and E, are independent conditioning on a set of variables X and returns a p-value for the null hypothesis. Thus, a small p-value (lower than the significance level) rejects the null hypothesis, indicating that the variables are not independent conditioning on a set of variables. If a p-value is high, the test fails to reject the null hypothesis, meaning that Y and E are likely independent conditioning on X. 

As an example, the \textit{Kernel conditional independence test} for $H_{m}$ ``$Merge Conflicts \CI Tests Volume  \mid$ $Commit Frequency$'' (i.e., $Merge Conflicts$ is independent of $Tests Volume$ conditioning on $Commit Frequency$) returns a p-value of 0.542159, failing to reject the null hypothesis for $H_{m}$. Thus, we assume that $Merge Conflicts$ and $Tests Volume$ are independent when conditioning on $Commit Frequency$ since the p-value is higher than a significance level of 0.05 (in fact, we fail to reject the null hypothesis).

\begin{comment}
\begin{table}[h]
	\caption{The results of the unconditional independence tests for RQ2.}
	\label{tab:hypothesis-testing-unconditional}
    \begin{tabular}{|l|l|}
		\hline
        \multicolumn{2}{|c|}{\textbf{Unconditional Independence}} \\ \hline   %& \multicolumn{1}{c|}{\textbf{Result}} \\ \hline
		\multicolumn{1}{|c|}{\textbf{Causal Hypotheses}}       & \multicolumn{1}{c|}{\textbf{R}} \\ \hline

		\begin{tabular}[c]{@{}l@{}}\textbf{H$_{1}$.} $Age\CI Commit Frequency$\end{tabular}                                                                                       & \textbf{49.5566*}                          \\ \hline
        \begin{tabular}[c]{@{}l@{}}\textbf{H$_{2}$.} $Age \CI Tests Volume$\end{tabular}         																			& \textbf{4.308763*}                          \\ \hline
        
		%\multicolumn{2}{l}{ $\dagger R > 1$}\\
        \multicolumn{2}{l}{*$ R > 1$}\\
        \multicolumn{2}{l}{$\CI$  ``Independent of...''}\\
	\end{tabular}
\end{table}
\end{comment}

\begin{table}[h]
	\caption{The results of the conditional independence tests for RQ2.}
	\label{tab:hypothesis-testing-conditional}
    \begin{tabular}{|l|l|}
		\hline
		\multicolumn{2}{|c|}{\textbf{Conditional Independence}} \\ \hline
		\multicolumn{1}{|c|}{\textbf{Causal Hypotheses}}       & \multicolumn{1}{c|}{\textbf{p-value}} \\ \hline   %& \multicolumn{1}{c|}{\textbf{Result}} \\ \hline
		\begin{tabular}[c]{@{}l@{}}\textbf{H$_{1}$.} $Age \CI Commit Frequency  \mid$ $Continuous Integration$\end{tabular}                                             & \textbf{1.41{e}{-07}*}                          \\ \hline
        \begin{tabular}[c]{@{}l@{}}\textbf{H$_{2}$.} $Age \CI Tests Volume \mid$ $Continuous Integration$\end{tabular}                                                  & \textbf{2.40{e}{-12}*}                          \\ \hline
        \begin{tabular}[c]{@{}l@{}}\textbf{H$_{3}$.} $Age \CI Communication \mid$ $Continuous Integration$\end{tabular}                                                & \textbf{1.16{e}{-06}*}                          \\ \hline
        \begin{tabular}[c]{@{}l@{}}\textbf{H$_{4}$.} $Age \CI Merge Conflicts  \mid$ $Continuous Integration$\end{tabular}                                              & 0.254925                          \\ \hline
        \begin{tabular}[c]{@{}l@{}}\textbf{H$_{5}$.} $Bug Report \CI Commit Frequency \mid$ $Communication$, $Continuous Integration$, $Tests Volume$\end{tabular}      & \textbf{0.000000*}                          \\ \hline
        \begin{tabular}[c]{@{}l@{}}\textbf{H$_{6}$.} $Bug Report\CI Merge Conflicts  \mid$ $Commit Frequency$, $Continuous Integration$\end{tabular}                    & 0.324586                          \\ \hline
        \begin{tabular}[c]{@{}l@{}}\textbf{H$_{7}$.} $Bug Report\CI Merge Conflicts  \mid$ $Communication$, $Tests Volume$,\\$Continuous Integration$\end{tabular}      & 0.344638                          \\ \hline
        \begin{tabular}[c]{@{}l@{}}\textbf{H$_{8}$.} $Communication \CI Merge Conflicts  \mid$ $Commit Frequency$, $Continuous Integration$\end{tabular}                & 0.486532                          \\ \hline
        \begin{tabular}[c]{@{}l@{}}\textbf{H$_{9}$.} $Communication \CI Tests Volume \mid$ $Continuous Integration$\end{tabular}                                        & \textbf{0.000000*}                          \\ \hline
        \begin{tabular}[c]{@{}l@{}}\textbf{H$_{10}$.}$Merge Conflicts \CI Tests Volume  \mid$ $Commit Frequency$, $Continuous Integration$\end{tabular}                 & 0.246583                          \\ \hline
        
		\multicolumn{2}{l}{ $*p < 0.05$}\\
        \multicolumn{2}{l}{$\CI$  ``Independent of...''}\\
        \multicolumn{2}{l}{$\mid$  ``Conditioning on...''}\\
	\end{tabular}
\end{table}

As we have seen in Table~\ref{tab:hypothesis-testing-conditional}, we reject hypotheses $H_{1}$, $H_{2}$, $H_{3}$, $H_{5}$, and $H_{9}$. These rejections are sufficient to conclude that if these d-separation implications from the causal DAG from our RQ1 (Fig.~\ref{fig:dag-rq1}) are not true, the causal DAG is not consistent with the empirical data. That means that the actual development processes from our studied projects (which generate the observed data) differ from the relationships shown in the literature-based DAG (Fig.~\ref{fig:dag-rq1}).

In simple terms, although our literature-based DAG indicates that $Age$ and $Commit Frequency$ are related only through $Continuous Integration$ influence (i.e., they are independent when conditioning on $Continuous Integration$), our tests have shown that they actually share a different relationship (i.e., the tests did not confirm the conditional independence). Similarly, our tests show that the relationship between $Age$ and the variables $Tests Volume$, $Communication$, and $Merge Conflicts$ are not independent when conditioning on $Continuous Integration$. Therefore, the tests suggest that $Age$ potentially has a direct relationship with these variables.

A parallel case is the one concerning $Communication$ and $Tests Volume$. The Literature-based DAG (Fig.~\ref{fig:dag-rq1}) presents a relationship between these two variables passing through $Continuous Integration$. However, our tests did not confirm the conditional independence $Communication \CI Tests Volume \mid$ $Continuous Integration$.

Another example is the relationship between $Bug Report$ and $Commit Frequency$. Even if we disregard the influences of $Age$, $Communication$, $Continuous Integration$, and $Tests Volume$, they remain dependent, which means that our data shows that $Bug Report$ and $Commit Frequency$ potentially have a direct relationship. Our data shows that $Bug Report$ potentially has a direct relationship with $Commit Frequency$.  In RQ3 (Section~\ref{rq3}), we deal with this set of observations on the failed hypothesis.

%\begin{tcolorbox}
%    RQ2: The theory-driven causal DAG from RQ1 is not consistent with the empirical data because the statistical implications from the DAG are not found in the dataset.
%\end{tcolorbox}

\subsection{RQ3. \RQthree\label{rq3}}

To discover a plausible causal structure, we start from the failed causal hypotheses in RQ2 and explore where the literature-based DAG (Section \ref{rq1}) does not match the evidence from the empirical dataset (Section \ref{rq2}). We propose an alternative DAG (data-validated DAG) based on the observed mismatches and the testing hypotheses derived from such mismatches (i.e., failed causal hypotheses).

%As defined in Section~\ref{method}, we build a new causal DAG through a hybrid approach mixing theory and data. To build the theory-data causal DAG, we conduct procedures in three steps: (i) Analyzing the rejected hypotheses in RQ2 and supported by d-separation rules (see Section~\ref{d-separation}) inferring changes in the structure of the DAG; (ii) Generating a new causal DAG from the identified changes; and (iii) We investigate the statistical implications from the new causal DAG.

%In the causal DAG in Fig.~\ref{fig:dag-rq1} there is a collider $Age \to Continuous Integration \leftarrow Merge Conflicts$, thus considering the d-separation rules (Section~\ref{d-separation}), we would expect an independence between $Age$ and $Merge Conflicts$. However, in the analysis from RQ2, we rejected hypothesis $H_{1}$ (Table~\ref{tab:hypothesis-testing}). Supposing the structure is not a collider when conditioning on $Continuous Integration$, $Age$ and $Merge Conflicts$ should become independent. We test this relationship with \textit{CondIndTest} and get a p-value of 0.1135986, confirming the independence between $Age$ and $Merge Conflicts$ when conditioning on $Continuous Integration$. Therefore, we invert the edge between $Continuous Integration$ and $Merge Conflicts$.
\subsubsection{The relationship between $Age$ and $Commit Frequency$ \label{section-age-commit}}
\begin{figure}[H]
    \centerline{\includegraphics[scale=.7]{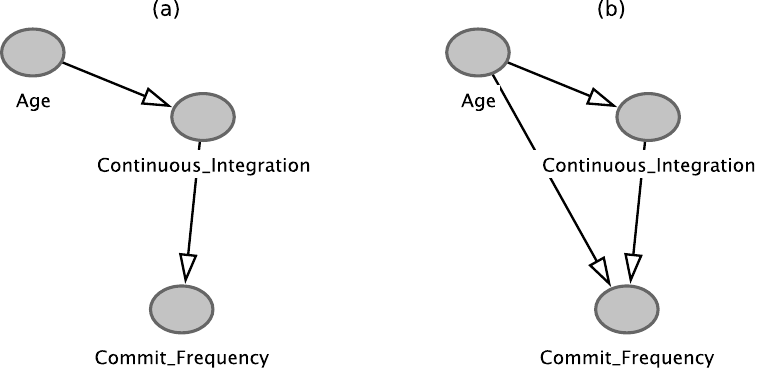}}
    \caption{(a) Causal structure (chain) involving $Age$, $Continuous Integration$, and $Commit Frequency$ as expressed in the literature-based causal DAG. (b) The new proposed causal structure concerning $Age$, $Continuous Integration$, and $Commit Frequency$ after statistical validations.}
    \label{fig:dag-rq3-h1}
\end{figure}

In the literature-based causal DAG in Fig.~\ref{fig:dag-rq1}, there is a chain $Age \to$ $Continuous Integration$ $\to Commit Frequency$, as shown in Fig.~\ref{fig:dag-rq3-h1} (a). Considering the d-separation rules (Section~\ref{d-separation}), we would expect that when conditioning on $Continuous Integration$, the association flow between $Age$ and $Commit Frequency$ would interrupt, so they should be independent. However, in the analysis from RQ2, we rejected hypothesis $H_{1}$ ($Age \CI Commit Frequency \mid$ $Continuous Integration$, see Table~\ref{tab:hypothesis-testing-conditional}). 

In this case, another path to the association between $Age$ and $Commit Frequency$ may exist. An alternative hypothesis is a wrong collider structure between $Age \to Bug Report \leftarrow Communication$. An association may occur through this path if this structure is not a collider. First, we tested the collider structure by formulating the hypothesis in Table~\ref{tab:hypothesis-testing-age-commits}. The hypothesis says that $Age$ would become independent of $Commit Frequency$ when conditioning on $Continuous Integration$ and $Bug Report$ ($Age \CI Commit Frequency \mid$ $Continuous Integration, Bug Report$). The result rejects such a hypothesis, which means a persistent dependence between $Age$ and $Commit Frequency$ when conditioning on $Continuous Integration$ and $Bug Report$. Therefore, we add a new edge between $Age$ and $Commit Frequency$, obtaining the structure shown in Fig.~\ref{fig:dag-rq3-h1} (b).

%Then, we test the unconditional independence between $Age$ and $Commit Frequency$, as expressed in Table~\ref{tab:hypothesis-testing-age-commits2}). We confirm this dependence between the variables and then add a new edge between $Age$ and $Commit Frequency$, obtaining the structure shown in Fig.~\ref{fig:dag-rq3-h1} (b).

\begin{table}[H]
	\caption{Conditional independence test for the relationship between $Age$ and $Commit Frequency$.}
	\label{tab:hypothesis-testing-age-commits}
    \begin{tabular}{|l|l|}
		\hline

		\multicolumn{2}{|c|}{\textbf{Conditional Independence}} \\ \hline
		\multicolumn{1}{|c|}{\textbf{Causal Hypotheses}}       & \multicolumn{1}{c|}{\textbf{p-value}} \\ \hline   %& \multicolumn{1}{c|}{\textbf{Result}} \\ \hline
		\begin{tabular}[c]{@{}l@{}}\textbf{H$_{1}$.} $Age \CI Commit Frequency \mid$ $Continuous Integration, Bug Report$\end{tabular}                & \textbf{$2.576492\mathrm{e}{-08}$*}                          \\ \hline
        \multicolumn{2}{l}{ $*p < 0.05 = dependence$}\\
        \multicolumn{2}{l}{$\CI$  ``Independent of...''}\\
        \multicolumn{2}{l}{$\mid$  ``Conditioning on...''}\\
	\end{tabular}
\end{table}

\begin{comment}
\begin{table}[h]
	\caption{Unconditional independence test for the relationship between $Age$ and $Commit Frequency$.}
	\label{tab:hypothesis-testing-age-commits2}
    \begin{tabular}{|l|l|}
		\hline
        \multicolumn{2}{|c|}{\textbf{Unconditional Independence}} \\ \hline   %& \multicolumn{1}{c|}{\textbf{Result}} \\ \hline
		\multicolumn{1}{|c|}{\textbf{Causal Hypotheses}}       & \multicolumn{1}{c|}{\textbf{R}} \\ \hline

		\begin{tabular}[c]{@{}l@{}}\textbf{H$_{1}$.} $Age\CI Commit Frequency$\end{tabular}                                                                                       & \textbf{49.5566*}                          \\ \hline        
		%\multicolumn{2}{l}{ $\dagger R > 1$}\\
        \multicolumn{2}{l}{*$ R > 1$}\\
        \multicolumn{2}{l}{$\CI$  ``Independent of...''}\\
	\end{tabular}
\end{table}
\end{comment}

\subsubsection{The relationship between $Age$ and $Tests Volume$ \label{section-age-tests}}
\begin{figure}[htbp]
    \centerline{\includegraphics[scale=.7]{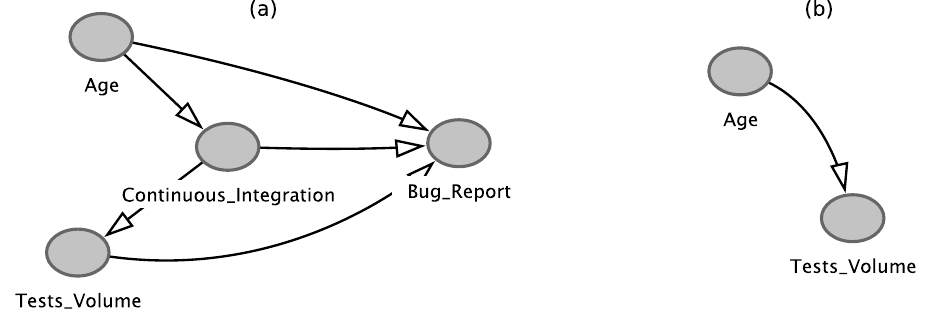}}
    \caption{(a) Causal structure involving $Age$ and $Tests Volume$ as expressed in the literature-based causal DAG. (b) The new proposed direct path concerning $Age$ and $Tests Volume$.}
    \label{fig:dag-rq3-h2}
\end{figure}

Analyzing hypothesis $H_{2}$ ($Age \CI Tests Volume \mid Continuous Integration$, see Table~\ref{tab:hypothesis-testing-conditional}) and the Fig.~\ref{fig:dag-rq3-h2} (a), we observe that given the following chain: $Age \to Continuous Integration \to Tests Volume$, there is an association between $Age$ and $Tests  Volume$, but this association would interrupt when conditioning on $Continuous Integration$. However, the test performed in RQ2 rejected such a hypothesis, revealing that $Age$ and $Tests Volume$ remain dependent. 

\begin{table}[h]
	\caption{Conditional independence tests for the relationship between $Age$ and $Tests Volume$.}
	\label{tab:hypothesis-testing-age-tests}
    \begin{tabular}{|l|l|}
		\hline

		\multicolumn{2}{|c|}{\textbf{Conditional Independence}} \\ \hline
		\multicolumn{1}{|c|}{\textbf{Causal Hypotheses}}       & \multicolumn{1}{c|}{\textbf{p-value}} \\ \hline   %& \multicolumn{1}{c|}{\textbf{Result}} \\ \hline
		\begin{tabular}[c]{@{}l@{}}\textbf{H$_{1}$.} $Age \CI Tests Volume \mid$ $Bug Report$\end{tabular}                & \textbf{0.0*}                          \\ \hline
        \begin{tabular}[c]{@{}l@{}}\textbf{H$_{2}$.} $Age \CI Tests Volume \mid$ $Continuous Integration, Bug Report$\end{tabular}    & \textbf{$8.518427\mathrm{e}{-10}$*}                          \\ \hline        
		\multicolumn{2}{l}{ $*p < 0.05$}\\
        \multicolumn{2}{l}{$\CI$  ``Independent of...''}\\
        \multicolumn{2}{l}{$\mid$  ``Conditioning on...''}\\
	\end{tabular}
\end{table}

Therefore, these variables should have a direct association, or the structure that links them through $Bug Report$ should be a chain rather than a collider. In this way, we should test this hypothesis in the form of conditional independence presented in Table~\ref{tab:hypothesis-testing-age-tests}, which expresses that if the structure is a chain when conditioning on the middle variable, $Age$, and $Tests Volume$ would become statistically independent. The test results in Table~\ref{tab:hypothesis-testing-age-tests} show that even conditioning on $CI, and Bug Report$, or just on $Bug Report$, the variables $Age$ and $Tests Volume$ remain dependent. Thus, we add a direct edge between them, as shown in Fig.~\ref{fig:dag-rq3-h2} (b).

\subsubsection{The relationship between $Age$ and $Communication$ \label{section-age-communication}}
\begin{figure}[htbp]
    \centerline{\includegraphics[scale=.7]{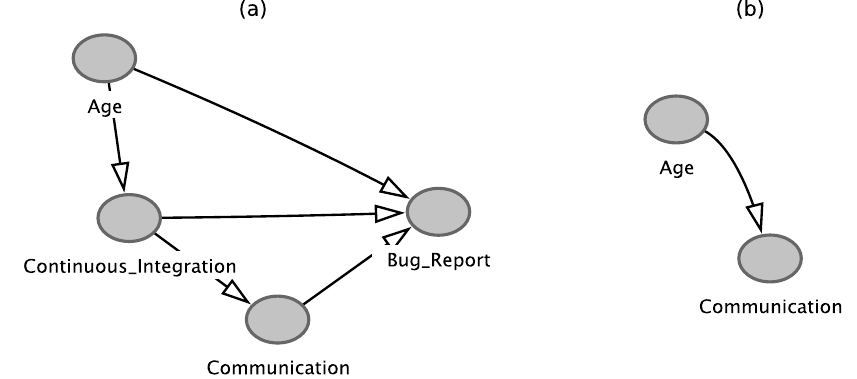}}
    \caption{(a) Causal paths between $Age$ and $Communication$ as expressed in the literature-based causal DAG. (b) The new proposed directed path between $Age$ and $Communication$.}
    \label{fig:dag-rq3-h4}
\end{figure}

Based on the existing paths between $Age$ and $Communication \mid Continuous Integration$ in the literature-based DAG (see Fig.~\ref{fig:dag-rq3-h4} (a)), the hypothesis $H_{3}$ ($Age \CI Communication  \mid$ $Continuous Integration$, see Table~\ref{tab:hypothesis-testing-conditional}) says that $Age$ and $Communication$ should be independent when conditioning on $Continuous Integration$. 

Rejecting hypothesis $H_{3}$ ($Age \CI Communication  \mid$ $Continuous Integration$) indicates that $Age$ and $Communication$ remain dependent. This dependency may be due to a wrong structure in the collider $Age \to Bug Report \leftarrow Communication$ that could be a chain, but in this case, when conditioning in $Bug Report$, $Age$, and $Communication$ would become independent.

Another hypothesis is a wrong structure in the chain $Age \to Continuous Integration \to Communication$ that may be a collider, and then when we conditioned in $Continuous Integration$, we opened the association flow, making $Age$ and $Communication$ dependent. In this way, we should test these hypotheses by testing the independencies depicted in Table~\ref{tab:hypothesis-testing-age-communication} and Table~\ref{tab:hypothesis-testing-age-communication2}.

\begin{table}[h]
	\caption{Conditional independence tests for the hypothesis related to relationship between $Age$ and $Communication$.}
	\label{tab:hypothesis-testing-age-communication}
    \begin{tabular}{|l|l|}
		\hline

		\multicolumn{2}{|c|}{\textbf{Conditional Independence}} \\ \hline
		\multicolumn{1}{|c|}{\textbf{Causal Hypotheses}}       & \multicolumn{1}{c|}{\textbf{p-value}} \\ \hline   %& \multicolumn{1}{c|}{\textbf{Result}} \\ \hline
		\begin{tabular}[c]{@{}l@{}}\textbf{H$_{1}$.} $Age \CI Communication \mid$ $Bug Report$\end{tabular}                & \textbf{0.0*}                          \\ \hline
        \begin{tabular}[c]{@{}l@{}}\textbf{H$_{1}$.} $Age \CI Communication \mid$ $Bug Report, Continuous Integration$\end{tabular}                & \textbf{1.251478{e}{-05}*}                          \\ \hline
		\multicolumn{2}{l}{ $*p < 0.05$}\\
        \multicolumn{2}{l}{$\CI$  ``Independent of...''}\\
        \multicolumn{2}{l}{$\mid$  ``Conditioning on...''}\\
	\end{tabular}
\end{table}

\begin{table}[h]
	\caption{Unconditional independence tests for the hypothesis related to relationship between $Age$ and $Communication$.}
	\label{tab:hypothesis-testing-age-communication2}
    \begin{tabular}{|l|l|}
		\hline
        \multicolumn{2}{|c|}{\textbf{Unconditional Independence}} \\ \hline   %& \multicolumn{1}{c|}{\textbf{Result}} \\ \hline
		\multicolumn{1}{|c|}{\textbf{Causal Hypotheses}}       & \multicolumn{1}{c|}{\textbf{R}} \\ \hline
        \begin{tabular}[c]{@{}l@{}}\textbf{H$_{2}$.} $Age \CI Communication$\end{tabular}       & \textbf{11.86268 *}                          \\ \hline
		%\multicolumn{2}{l}{ $\dagger R > 1$}\\
        \multicolumn{2}{l}{*$ R > 1$}\\
        \multicolumn{2}{l}{$\CI$  ``Independent of...''}\\
	\end{tabular}
\end{table}

The test result in Table~\ref {tab:hypothesis-testing-age-communication} shows that even conditioning on $Bug Report$, the variables $Age$ and $Communication$ remain dependent. Thus, there exists a dependency through another path. The result in Table~\ref{tab:hypothesis-testing-age-communication2} does not confirm the hypothesis of a collider in $Age \to Continuous Integration \leftarrow Communication$ since without conditioning on $Continuous Integration$, $Age$ and $Communication$ remain dependent. Therefore, we add an edge $Age\to Communication$, as illustrated in Fig.~\ref{fig:dag-rq3-h4} (b).

\subsubsection{The relationship between $Commit Frequency$ and $Bug Report$ \label{section-commit-bug}}
\begin{figure}[htbp]
    \centerline{\includegraphics[scale=.7]{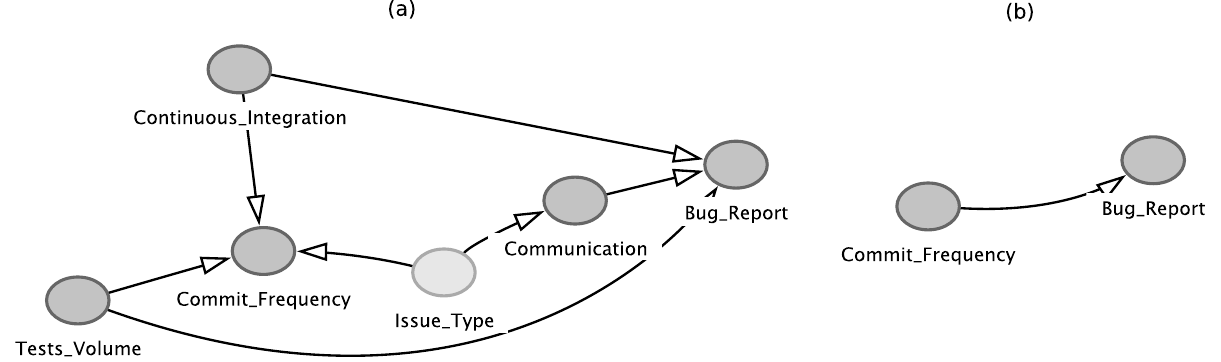}}
    \caption{(a) Causal paths between $Commit Frequency$ and $Bug Reports$ as expressed in the literature-based causal DAG. (b) The new proposed directed path between $Commit Frequency$ and $Bug Reports$.}
    \label{fig:dag-rq3-h5}
\end{figure}

The rejection of hypothesis $H_{5}$ ($Bug Report \CI Commit Frequency \mid$ $Communication$, $Continuous Integration$, $Tests Volume$, see Table~\ref{tab:hypothesis-testing-conditional}), similarly, indicates that the structures shown in Fig.~\ref{fig:dag-rq3-h5} (a) are not accurate. 

According to these structures, we expected that $Bug Report$ and $Commit Frequency$ would become independent when conditioning on $Continuous Integration$, $Communication$, and $Tests Volume$. However, the test did not confirm the independence (see Table~\ref{tab:hypothesis-testing-conditional}), even testing all variables conditioned together or separately. On the other hand, we tested if $Bug Report$ and $Commit Frequency$ are unconditionally independent and got an $R-value$ of 6.009902, indicating a dependency between them.

These tests suggest the existence of another path between $Bug Report$ and $Commit Frequency$. Thus, we add a new edge linking them as shown in Fig.~\ref{fig:dag-rq3-h5} (b).

\subsubsection{The relationship between $Communication$ and $Tests Volume$\label{section-ci-tests}}
\begin{figure}[h]
    \centerline{\includegraphics[scale=.55]{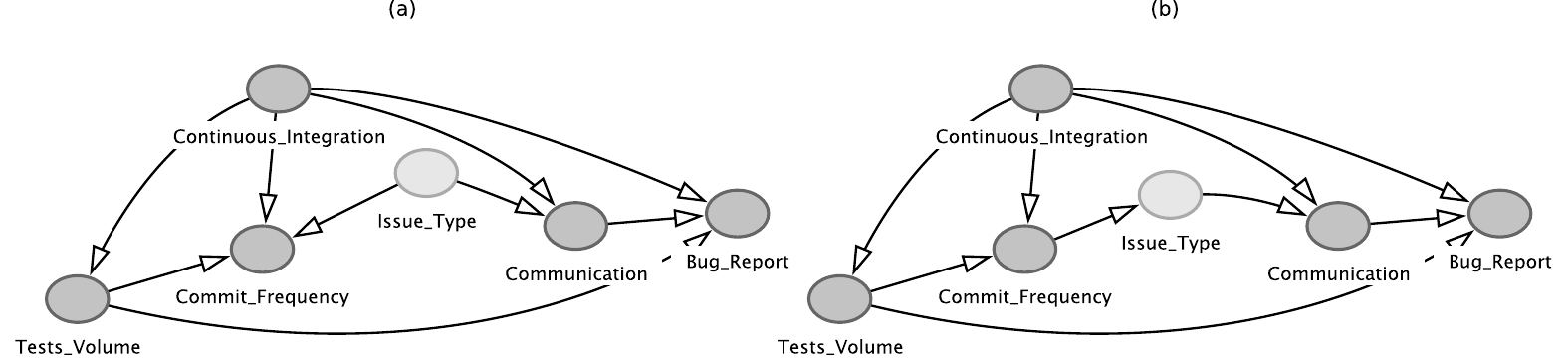}}
    \caption{(a) Causal structure between $Communication$, and $Tests Volume$ as expressed in the literature-based causal DAG. (b) The new proposed structure between $Communication$ and $Tests Volume$ after statistical validations.}
    \label{fig:dag-rq3-h9}
\end{figure}

Analyzing the rejection of hypothesis $H_{9}$, we observed that when conditioning $Communication$ and $Tests Volume$ on $Continuous Integration$ (see Fig.~\ref{fig:dag-rq3-h9} (a)), the dependence between $Communication$ and $Tests Volume$ does not disappear. We then tested if the collider $Tests Volume \to Commit Frequency \leftarrow Issue Type$ would be wrong. If this hypothesis is correct, the association is flowing through this path, then conditioning on $Commit Frequency$ would block the association.

In this sense, we test the conditional independence presented in Table~\ref{tab:hypothesis-testing-communication-tests} and confirm the hypothesis that the structure is not a collider. Therefore, we reorient the edge between $Commit Frequency$ and $Issue Type$, as present Fig.~\ref{fig:dag-rq3-h9} (b).

\begin{table}[h]
	\caption{Conditional independence tests for the hypothesis related to relationship between $Communication$ and $Tests Volume$.}
	\label{tab:hypothesis-testing-communication-tests}
    \begin{tabular}{|l|l|}
		\hline

		\multicolumn{2}{|c|}{\textbf{Conditional Independence}} \\ \hline
		\multicolumn{1}{|c|}{\textbf{Causal Hypotheses}}       & \multicolumn{1}{c|}{\textbf{p-value}} \\ \hline   %& \multicolumn{1}{c|}{\textbf{Result}} \\ \hline
		\begin{tabular}[c]{@{}l@{}}\textbf{H$_{1}$.}  $Communication \CI$ $Tests Volume \mid$ $Continuous Integration$, $Commit Frequency$\end{tabular}                & 0.862325                          \\ \hline
		\multicolumn{2}{l}{ $*p < 0.05$}\\
        \multicolumn{2}{l}{$\CI$  ``Independent of...''}\\
        \multicolumn{2}{l}{$\mid$  ``Conditioning on...''}\\
	\end{tabular}
\end{table}

\subsubsection{Data-Validated Causal DAG \label{section-data-validated-dag}}
                                                
\begin{figure}[h]
    \centerline{\includegraphics[scale=.55]{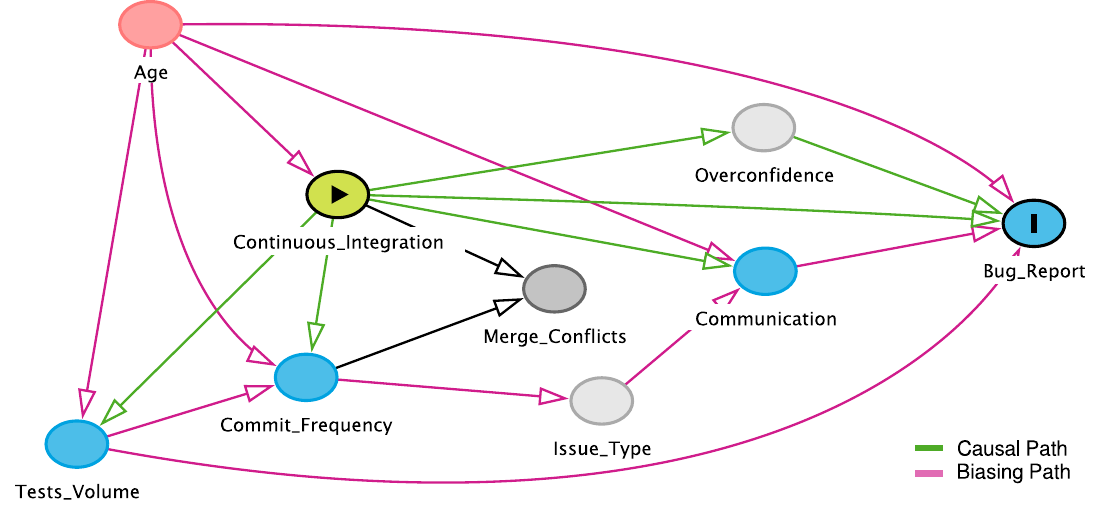}}
    \caption{Initial version of data-validated DAG for CI, Bug Reports and their co-variables.}
    \label{fig:dag-rq3_0}
\end{figure}

Afterward, we obtain the first version of data-validated DAG in Fig.~\ref{fig:dag-rq3_0}, raising a new set of statistical implications to test. Table~\ref{tab:rq3-hypothesis-testing-v0} shows such implications as causal hypotheses and the results of the independence tests. 

\begin{table}[H]
	\caption{The results of the conditional independence tests for RQ3.}
	\label{tab:rq3-hypothesis-testing-v0}
    \begin{tabular}{|l|l|}
		\hline
		\multicolumn{2}{|c|}{\textbf{Conditional Independence}} \\ \hline
		\multicolumn{1}{|c|}{\textbf{Causal Hypotheses}}       & \multicolumn{1}{c|}{\textbf{p-value}} \\ \hline   %& \multicolumn{1}{c|}{\textbf{Result}} \\ \hline
		\begin{tabular}[c]{@{}l@{}}\textbf{H$_{1}$.} $Age \CI Merge Conflicts  \mid$ $Commit Frequency, Continuous Integration$\end{tabular}                    & 0.269635                          \\ \hline
        \begin{tabular}[c]{@{}l@{}}\textbf{H$_{2}$.} $Bug Report \CI Merge Conflicts  \mid$ $Commit Frequency$, $Continuous Integration$\end{tabular}           & 0.321338                         \\ \hline
        \begin{tabular}[c]{@{}l@{}}\textbf{H$_{3}$.} $Communication \CI Merge Conflicts \mid$ $Commit Frequency$, $Continuous Integration$\end{tabular}         & 0.486532                          \\ \hline
        \begin{tabular}[c]{@{}l@{}}\textbf{H$_{4}$.} $Communication \CI Tests Volume \mid$ $Age, Commit Frequency, Continuous Integration$\end{tabular}         & \textbf{2.05{e}{-12} *}                          \\ \hline
        \begin{tabular}[c]{@{}l@{}}\textbf{H$_{5}$.} $Merge Conflicts \CI Tests Volume \mid$ $Commit Frequency$, $Continuous Integration$\end{tabular}          & 0.246583                          \\ \hline

        \multicolumn{2}{l}{*$p < 0.05$}\\
        \multicolumn{2}{l}{$\CI$  ``Independent of...''}\\
        \multicolumn{2}{l}{$\mid$  ``Conditioning on...''}\\
	\end{tabular}
\end{table}

\begin{figure}[htbp]
    \centerline{\includegraphics[scale=.55]{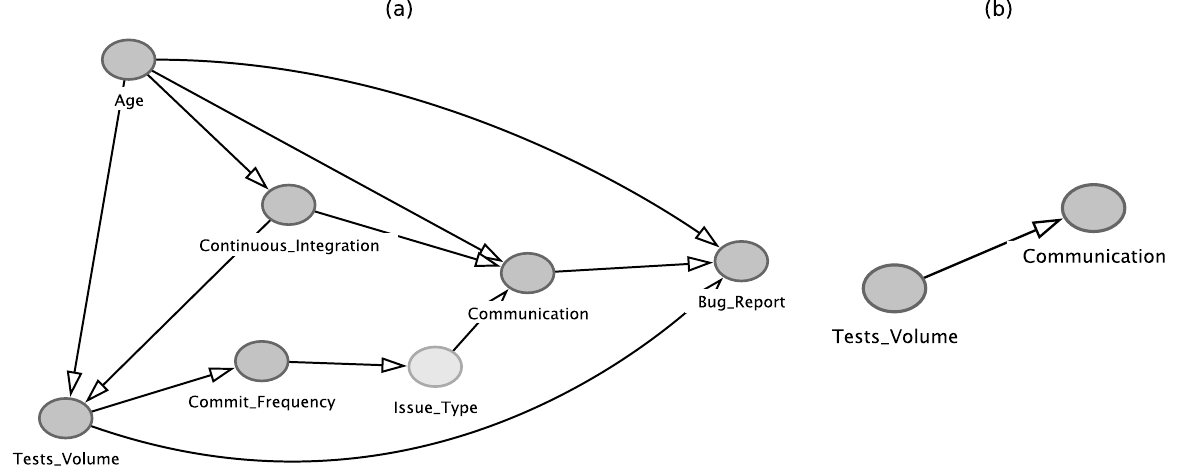}}
    \caption{(a) Causal paths between $Tests Volume$ and $Communication$ as expressed in the literature-based causal DAG. (b) The new proposed direct path between $Tests Volume$ and $Communication$.}
    \label{fig:dag-rq3-h4b}
\end{figure}

\begin{figure}[htbp]
    \centerline{\includegraphics[scale=.55]{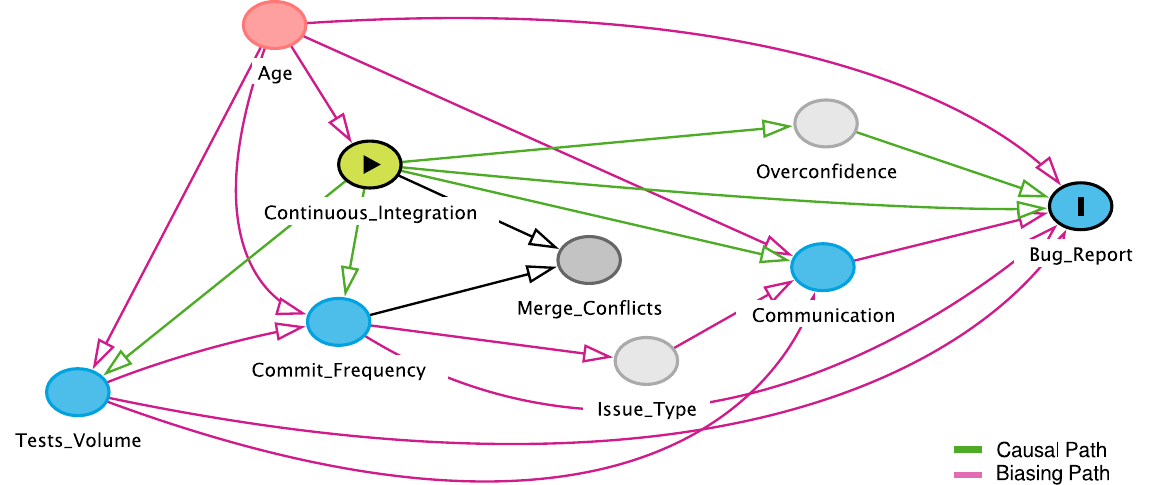}}
    \caption{data-validated DAG for CI, Bug Reports and their co-variables.}
    \label{fig:dag-rq3}
\end{figure}

After the statistical analysis, the $H_{4}$ ($Communication \CI Tests Volume \mid$ $Age$, $Commit Frequency$, $Continuous Integration$, see Table~\ref{tab:rq3-hypothesis-testing-v0}) was rejected. This result means that even conditioning on $Age, Commit Frequency$ and $Continuous Integration$, the variables $Communication$ and $Tests Volume$ remain dependent, i.e., another open path exists between them. Since the paths by $Bug Report$ is a collider (see Fig.~\ref{fig:dag-rq3-h4b} (a)), we may infer that there is a path that this first version of data-validated DAG (Fig.~\ref{fig:dag-rq3_0}) does not reflect the true causal relationship. Therefore, we add a new edge between $Tests Volume$ and $Communication$ (as shown in Fig.~\ref{fig:dag-rq3-h4b} (b)) and achieve a new version of the data-validated DAG shown in Fig.~\ref{fig:dag-rq3}.

The data-validated DAG shown in Fig.~\ref{fig:dag-rq3} raises a new set of statistical implications depicted in Table~\ref{tab:rq3-hypothesis-testing-v1}. After the statistical analysis, all these four hypotheses were confirmed, meaning that the dataset upholds all implications of the data-validated DAG. Thus, we can conclude that we find a plausible model in our data-validated DAG (Fig.~\ref{fig:dag-rq3}).

\begin{table}[H]
	\caption{The results of the conditional independence tests for RQ3.}
	\label{tab:rq3-hypothesis-testing-v1}
    \begin{tabular}{|l|l|}
		\hline
		\multicolumn{2}{|c|}{\textbf{Conditional Independence}} \\ \hline
		\multicolumn{1}{|c|}{\textbf{Causal Hypotheses}}       & \multicolumn{1}{c|}{\textbf{p-value}} \\ \hline   %& \multicolumn{1}{c|}{\textbf{Result}} \\ \hline
		\begin{tabular}[c]{@{}l@{}}\textbf{H$_{1}$.} $Age \CI Merge Conflicts  \mid$ $Commit Frequency$, $Continuous Integration$\end{tabular}                  & 0.269635                          \\ \hline
        \begin{tabular}[c]{@{}l@{}}\textbf{H$_{2}$.} $Bug Report \CI Merge Conflicts  \mid$ $Commit Frequency$, $Continuous Integration$\end{tabular}           & 0.321338                         \\ \hline
        \begin{tabular}[c]{@{}l@{}}\textbf{H$_{3}$.} $Communication \CI Merge Conflicts \mid$ $Commit Frequency$, $Continuous Integration$\end{tabular}         & 0.486532                          \\ \hline
        \begin{tabular}[c]{@{}l@{}}\textbf{H$_{5}$.} $Merge Conflicts \CI Tests Volume \mid$ $Commit Frequency$, $Continuous Integration$\end{tabular}          & 0.384515                          \\ \hline
        
        \multicolumn{2}{l}{*$p < 0.05$}\\
        \multicolumn{2}{l}{$\CI$  ``Independent of...''}\\
        \multicolumn{2}{l}{$\mid$  ``Conditioning on...''}\\
	\end{tabular}
\end{table}

\section{Discussion \label{discussion}}

In RQ1~\ref{rq1}, we cataloged a set of claims related to Continuous Integration that we verified empirically in RQ2~\ref{rq2}. Finally, in the RQ3~\ref{rq3}, we proposed a new causal DAG mixing the literature knowledge with the empirical data. 

The data-validated DAG shown in Fig.~\ref{fig:dag-rq3} shows a more significant influence of $Age$ than in the previous literature-based DAG (Fig.~\ref{fig:dag-rq1}). $Age$ is an important source of confounding effects since it relates to $Tests Volume$, $Commit Frequency$, $Continuous Integration$, $Communication$, and $Bug Report$. 

In the data-validated DAG, we can also observe that $Continuous Integration$ influences $Commit Frequency$ and $Merge Conflicts$, while $Commit Frequency$ also contributes to more $Merge Conflicts$ and indirectly to more $communication$. This way, $Continuous Integration$ influences, directly and indirectly, to obtain more $Communication$ in the software development process. $Communication$, in turn, has a causal relationship with $Bug Report$. 

The negative side of $Continuous Integration$ is the occurrence of $Merge Conflicts$ and the promotion of an $Overconfidence$ in the development environment. We conjecture that CI is believed to decrease the commit size because of the increased commit frequency. This higher frequency engenders a favorable scenario for an increase in merge conflicts.

As for the overconfidence, it may arise from the belief that the automated environment provided by CI is always reliable. Thus, we refer to it as confidence over the reasonable, which is not necessarily grounded on objective mechanisms and metrics that give such security and confidence(e.g., code coverage, test quality, code smells checking). Such an $Overconfidence$ may increase $Bug Report$ since developers may relax their attention to detail, outsourcing such quality control to automation developed in the scope of CI. 

Finally, we observe that $Continuous Integration$ also, directly and indirectly, influences $Bug Report$. $Continuous Integration$ has a direct relationship with $Bug Report$ and has an indirect influence through $Communication$, $Commit Frequency$, $Tests Volume$, and $Overconfidence$. While $Communication$, $Commit Frequency$, and $Tests Volume$ affects $Bug Report$ negatively, $Overconfidence$ affects $Bug Report$ positively.

%Such a set of observations raises implications for researchers and practitioners.

\subsection{Implications for researchers}
The DAGs presented in this study, especially the data-validated DAG (Fig.~\ref{fig:dag-rq3}), may serve as a baseline for further CI and software quality investigations. This DAG brings findings from the combination of assumptions from the literature confronted with an empirical dataset. Other studies can reinforce, refute, or expand our findings and the DAGs, contributing to a more robust theory on the effects of CI on software quality (our replication package\footnote{\replication} may be helpful). 

We can also consider the differences emerging from the comparison between the literature-based DAG (Fig.~\ref{fig:dag-rq1}) and the data-validated DAG (Fig.~\ref{fig:dag-rq3}). The changed associations arise from the data and reveal exciting findings to be further investigated by researchers. As well as the edges additions originated in data-validated DAG.

%The research community can undertake efforts to deepen the understanding of these phenomena, i.e., the interactions between Continuous Integration and Commit patterns. Studies may investigate, for example, whether CI is associated with certain types or patterns of merge conflicts. Alternatively, they can study how developers could avoid merge conflicts in a continuous environment with frequent commits.

%h1, h2, h3, h5
%h1
The data-validated DAG (Fig.~\ref{fig:dag-rq3}) added new edges. The addition of $Age \to Commit Frequency$ (Section~\ref{section-age-commit}) aligns with  Zhao et al.~\cite{zhao2017impact} that found a slight decreasing trend in the number of non-merge commits over time. As interpreted by Brindescu et al.~\cite{brindescu2014}, this finding could be related to the project activity level changes during their lifecycle. The projects tend to switch activities from adding new features to performing corrective changes and similar ones~\cite{brindescu2014}.

%h2
In this line, as the project ages, more $Tests Volume$ tend to exist ($Age \to Test Volume$). Despite the literature used to build literature-based DAG in our RQ1 (Section \ref{rq1}) does not explore the direct relationship between $Age$ and $Tests Volume$, our dataset analysis reveals it. Sizilio Nery et al.~\cite{sizilio2019empirical} investigates the evolution of test ratio (a metric similar to our $Tests Volume$) and relates a change in the test ratio evolution trend to CI employment, NOCI projects have a negligible change in the test ratio over time. In our dataset, a causal relationship exists between $Age$ and $Tests Volume$, i.e., as the $Age$ of a project advances, the greater the $Tests Volume$ independently of CI employment or not. Future investigations may study qualitatively how the tests evolve regarding the quality and motivation behind such test efforts. 

%h3
So is the $Age$ influence on $Commit Frequency$ and $Tests Volume$, $Age$ also causes an increase in $Communication$, in terms of pull requests comments.
For example, an interesting observation arises from the causal flow $Age \to Communication$ of our data-validated DAG. We note that in the literature, Cassee et al. [46] investigate the association between CI adoption and pull-request communication. Our study, however, identifies an effect on communication separated from CI, indicating that the Age of a project may affect $Communication$ over time rather than CI. Therefore,  our data-validated DAG can be used to better understand the existing associations observed in the literature.

%there is room for further studies on other forms and channels of communication and how $Age$ influence it, e.g., is this $Communication$ increasing related to commit frequency? Researchers can also conduct a study to collect the developers' perceptions of the impact of Age and CI on communication.

In this line of thought, we demonstrate in Section~\ref{section-data-validated-dag} that $Tests Volume$ also influences $Communication$ ($Tests Volume \to Communication$). It is possible that as test volume increases, more discussion is generated regarding the commits. There is an interrelationship between $Tests Volume$, $Commit Frequency$, and $Communication$, as discussed by Fowler: ``Integration is about communication''~\cite{fowler_ci}. 

Similarly, these three variables affect $Bug Report$. Section~\ref{section-commit-bug} presents an addition of the edge $Commit Frequency \to Bug Report$, suggesting that the more frequent the commits, the less the $Bug Reports$. Indeed, the more integration, the more builds and quality checks are performed. Eyolfson et al.~\cite{Eyolfson2013} found that more frequent committers produce fewer bug-introducing commits. $Commit Frequency$ could have a similar foundation: the more frequent the commits, the more knowledge, and familiarity with the domain, architecture, and tools used in the project, providing less faulty development. Researchers could perform empirical studies on the relationship between commit frequency and code or project quality.

If more tests and frequent commits mean more communication, then CI is essentially promoting more communication, directly and indirectly. Moreover, more communication generates more knowledge and fewer errors, thus producing a high-quality software product. Cassee et al.~\cite{cassee2020silent} highlight the necessity of investigating the social aspects of software development. In their work, Cassee et al.~\cite{cassee2020silent} relate CI with a decrease in the number of comments in code reviews; however, they call CI a ``Silent Helper'' because CI plays a role communicating sufficiently to maintain the same level of activity of projects with more comments. This view corroborates the idea that the CI environment, $Commit Frequency$, $Tests Volume$, all of it is about communication.

Further studies may investigate this techno-social aspect, answering how and why $Commit Frequency$ and $Tests Volume$ can potentialize communication. Furthermore, investigating how the CI environment contributes to communication, i.e., in which aspects it communicates and the mechanisms, tools, and metrics through which CI is manifested, may mitigate the harmful false sense of confidence ($Overconfidence$) generated by CI in some environments. For instance, Bernardo et al.~\cite{bernardo2023} highlight CI influencing higher confidence during code review. 

In addition, future studies may analyze whether CI is the cause or the consequence for attracting or attaining more experienced (core) developers. If a clear causal flow exists between $Experienced Developers \to CI$ or $CI \to Experienced Developers$, other potential investigations would open, such as investigating whether CI generates more or less bug reports because, for instance, experienced developers detect more bugs.

The data-validated causal DAG (Fig.~\ref{fig:dag-rq3}) shows $Age$ as a significant source of confounding (constituting a backdoor path) for $Continuous Integration$ and $Bug Report$. Thus, researchers should be more attentive to controlling for $Age$ when studying the effects of CI. Additionally, researchers may investigate the rationale behind this relationship between older projects and CI projects. They may explore, for example, the decision process for CI adoption and the challenges involved in such a process that could postpone the adoption.

\subsection{Practical implications}

Our study and the data-validated DAG (Fig.~\ref{fig:dag-rq3}) raise practical implications. Software developers may understand that they potentially decrease the bug reports when employing CI in their projects. CI causes an effect on the $Bug Report$, encourages team $Communication$, and increases the $Commit Frequency$ (in turn, both positively affect bug reports). Herefore, considering the observed literature-reported relationships and the variables we were able to measure in our study, CI has a causal effect on the number of Bug Reports, which, in turn, is our proxy for software quality. 

On the other hand, CI projects have greater values for the $Age$ variable. Adopting CI may be related to greater project maturity. This demand may be related to the difficulties of implementing CI and a demand for more qualified and experienced professionals. Awareness of (over)confidence in a flaky automated environment is also essential. Investing in building a reliable environment could mitigate the negative overconfidence effect.

%\subsection{RQ3. What are the qualitative differences on Bug reports between CI and No CI projects?\label{rq3}}

%\subsubsection{Motivation}
%\dots\\
%\subsubsection{Approach}
%\dots\\
%\subsubsection{Results}
%\dots

\section{Threats To Validity\label{threats}}
In this section, we discuss the threats to the validity of our study.

\textbf{Construct Validity. } 
There are construct threats associated with the searching for associations between CI and software quality from existing software engineering literature. These associations are used to build the DAG showing how CI may influence software quality with potential confounding variables (RQ1). To reduce the threats associated with the search for such associations, we have mainly used existing systematic literature reviews and empirical studies to search for research work associated with CI and software quality. However, there are always associations (open questions) that are not captured by existing studies or even studies that were not found and considered in our analysis. In addition to that, we have also considered the associations related to CI sub-practices that are present in the selected papers as a direct relationship between CI and other important software development aspects.  

We also have construct threats associated with the repository mining of the collected data that supports the statistical tests of the associations in the DAG (RQ2). We used GitHub and Travis CI APIs to collect data about the selected projects and compute the metrics. Any bias in how we compute such metrics can affect our results. For example, we can measure communication in other ways, such as calls, meetings, emails, and others. As another example, merge conflicts and bug issues were cataloged based on heuristics that could carry biases preventing our scripts from precisely identifying all issues and conflicts.
However, we used our best efforts to collect the seven considered metrics in this study. We seek to follow methods and metrics already known in the software engineering community, as we show throughout Section~\ref{method}.

\textbf{Internal Validity. }
To compute the metrics associated with our dataset, we aggregate them into months. Such a strategy can produce a lack of information (compared to a week or day-based analysis) associated with the metrics of interest. Recent empirical studies~\cite{sizilio2019empirical,bernardo2018studying} that investigate the impact of CI have adopted a similar month-based strategy.

\textbf{External Validity. } External threats concern the extent to which we can generalize our results. Our work analyzed a total of 59,034 commits, 16,619 builds, and 3,199 bug reports of 148 Github open-source projects - 74 CI and 74 no-CI projects. Nevertheless, we collected a consistent dataset representative of the open-source community, which validates our DAG. However, other studies would be needed to reinforce, refute, or modify our causal DAG. The more projects and metrics reliable and consistently collected, the more robust would be the causal DAG and the causal theory proposed. This study is not intended to be definitive concerning the causal relationship between CI and software quality but rather to be a starting point in such a direction. More studies are needed to evolve our understanding of causal relationships between CI and quality.

\section{Conclusion\label{conclusion}}
Continuous Integration (CI) has increased in popularity in the last decades with the expectation of providing several software development benefits. One of these benefits associated with CI is increased software quality. This study expands the current software engineering body of knowledge by investigating the association between CI and software quality from a causal perspective. We reviewed the literature cataloging relevant CI and software quality associations and adjacent relationships. Then we built the literature-based causal DAG (RQ1) expressing these associations. We mined software repositories to empirically assess the literature-based causal DAG (RQ2), analyzing releases in 12 activity months from 148 software projects.

%The literature knowledge mapped into our causal DAG points out a causal effect of CI in bug reports, and that effect also has a contribution from community communication, as well as the effect has a negative contribution from developers' overconfidence. We should be aware of test volume and project age influences that, according to literature assumptions, may confound the CI effect.

Analyzing the testable implications from the literature-based causal DAG generated from literature assumptions, we found in RQ2 some associations that do not exist in the collected dataset. Then, relying on a hybrid literature-data approach, we proposed a new data-validated causal DAG (RQ3), expressing the relationship between the relevant variables. 

We discussed research opportunities from the differences cataloged between literature-based DAG and the data-validated causal DAG. Researchers and practitioners can benefit from our findings based on the proposed data-validated causal DAG that, in summary, shows a positive causal effect from CI on bug reports. CI also influences developers' communication, reinforcing bug reports benefits. The developers' overconfidence is still a concern in CI environments and could negatively affect bug reports.

%Other studies may complement, reinforce or propose new causal DAGs through larger datasets, considering 

%\section*{Acknowledgment}

\begin{acknowledgements}
	This work is partially supported by INES (\url{www.ines.org.br}), CNPq grants 465614/2014-0 and 425211/2018-5, CAPES grant 88887.136410/2017- 00, and FACEPE grants APQ-0399-1.03/17 and PRONEX APQ/0388-1.03/14.
\end{acknowledgements}

  \textbf{Data Availability} 
  All data is available via an online appendix: \url{\replication}.
  
  \section*{Declarations}
  \textbf{Conflict of interests} 
  The authors declare that they have no conflict of interest.

\vspace{12pt}

\newpage

\end{document}